%% file: SUS-16-036_temp.tex
\pdfoutput=1

\documentclass[11pt,twoside,a4paper,cmspaper,final,collab]{cms-tdr}
\def\svnVersion{433057P}\def\cmsCernNoTag{CERN-EP-2017-084}\def\cmsCernDate{\today}\def\cmsMessage{Published in the European Physical Journal C as \href{http://dx.doi.org/10.1140/epjc/s10052-017-5267-x}{\doi{10.1140/epjc/s10052-017-5267-x.}}}

\begin{document}\cmsNoteHeader{SUS-16-036}

\hyphenation{had-ron-i-za-tion}
\hyphenation{cal-or-i-me-ter}
\hyphenation{de-vices}

\RCS$Revision: 430382 $
\RCS$HeadURL: svn+ssh://svn.cern.ch/reps/tdr2/papers/SUS-16-036/trunk/SUS-16-036.tex $
\RCS$Id: SUS-16-036.tex 430382 2017-10-21 15:43:32Z fgolf $
\newlength\cmsFigWidth
\ifthenelse{\boolean{cms@external}}{\setlength\cmsFigWidth{0.40\textwidth}}{\setlength\cmsFigWidth{0.49\textwidth}}
\ifthenelse{\boolean{cms@external}}{\providecommand{\cmsLeft}{upper\xspace}}{\providecommand{\cmsLeft}{left\xspace}}
\ifthenelse{\boolean{cms@external}}{\providecommand{\cmsRight}{lower\xspace}}{\providecommand{\cmsRight}{right\xspace}}
\ifthenelse{\boolean{cms@external}}{\providecommand{\cmsLLeft}{Upper\xspace}}{\providecommand{\cmsLLeft}{Left\xspace}}
\ifthenelse{\boolean{cms@external}}{\providecommand{\cmsRRight}{Lower\xspace}}{\providecommand{\cmsRRight}{Right\xspace}}
\ifthenelse{\boolean{cms@external}}{\providecommand{\cmsMiddle}{Lower\xspace}}{\providecommand{\cmsMiddle}{Right\xspace}}
\providecommand{\PSGcmpDo}{\ensuremath{\widetilde{\chi}^\mp_{1}}\xspace}

\cmsNoteHeader{SUS-16-036}
\title{Search for new phenomena with the \texorpdfstring{\mttwo}{MT2} variable in the all-hadronic final state produced in proton-proton collisions at \texorpdfstring{$\sqrt{s} = 13$\TeV}{sqrt(s) = 13 TeV}}
\titlerunning{Search for new phenomena with \mttwo at 13\TeV}

\date{\today}

\newcommand{\Lint}{35.9\fbinv\xspace}
\newcommand{\mttwo}{\ensuremath{M_{\mathrm{T2}}}\xspace}
\newcommand{\Ht}{\ensuremath{H_{\mathrm{T}}}\xspace}
\newcommand{\Met}{\ptmiss}
\newcommand{\Mht}{\ensuremath{H_{\mathrm{T}}^{\mathrm{miss}}}\xspace}
\newcommand{\vMet}{\ptvecmiss}
\newcommand{\vMht}{\ensuremath{\vec{H}_{\mathrm{T}}^{\text{miss}}}\xspace}
\newcommand{\dpmin}{\ensuremath{\Delta\phi_{\text{min}}}\xspace}
\newcommand{\njets}{\ensuremath{N_{\mathrm{j}}}\xspace}
\newcommand{\nbtags}{\ensuremath{N_{\PQb}}\xspace}
\newcommand{\Mt}{\ensuremath{M_{\mathrm{T}}}\xspace}
\newcommand{\wjets}{\ensuremath{\PW\text{+jets}}\xspace}
\newcommand{\zjets}{\ensuremath{\PZ\text{+jets}}\xspace}
\newcommand{\vjets}{\ensuremath{\text{V}\text{+jets}} \ensuremath{(\text{V}=\PZ,\PW)}\xspace}
\newcommand{\ttjets}{\ensuremath{\ttbar \text{+jets}}\xspace}
\newcommand{\ttV}{\ensuremath{\ttbar\text{V} (\text{V}=\PZ,\PW)}\xspace}
\newcommand{\znunu}{\ensuremath{\PZ\to\PGn\PAGn}\xspace}
\newcommand{\ptj}{\ensuremath{\pt^{\text{jet1}}}\xspace}
\newcommand{\gluino}{\PSg}
\newcommand{\lsp}{\PSGczDo}
\newcommand{\astop}{\PASQt}
\providecommand{\NA}{\text{---}\xspace}

\abstract{
A search for new phenomena is performed using events with jets and significant transverse momentum imbalance, as inferred through the \mttwo variable.
The results are based on a sample of proton-proton collisions
collected in 2016 at a center-of-mass energy of 13\TeV with the CMS detector
and corresponding to an integrated luminosity of 35.9\fbinv.
No excess event yield is observed above the predicted standard model background,
and the results are interpreted as
exclusion limits at 95\% confidence level
on the masses of predicted particles in a variety of simplified models
of $R$-parity conserving supersymmetry.  Depending on the details of the model, 95\% confidence level lower
limits on  the gluino (light-flavor squark) masses are placed up to 2025 (1550)\GeV.
Mass limits as high as 1070 (1175)\GeV are set on the masses of top (bottom) squarks.
Information is provided to enable re-interpretation of these results,
including model-independent limits on the number of non-standard model events for a set of simplified, inclusive search regions.
}

\hypersetup{
pdfauthor={CMS Collaboration},
pdftitle={Search for new phenomena with the MT2 variable in the all-hadronic final state produced in proton-proton collisions at sqrt(s) = 13 TeV},
pdfsubject={CMS},
pdfkeywords={CMS, physics, supersymmetry}}

\maketitle

\section{Introduction}
\label{sec:intro}

We present results of a search for new phenomena in events with jets and significant transverse momentum imbalance in
proton-proton collisions at $\sqrt{s} = 13\TeV$.
Such searches were previously conducted by both the ATLAS~\cite{ATLASat13TeV:manyjets,ATLASat13TeV:monojet,ATLASat13TeV:hadronic,ATLASat13TeV:manyb,ATLASat13TeV:2b}
and CMS~\cite{MT2at13TeV,RA2at13TeV,RAZORat13TeV,AlphaTat13TeV} Collaborations.
Our search builds on the work presented in Ref.~\cite{MT2at13TeV}, using improved methods to estimate the background from standard model (SM) processes and a data set corresponding to an integrated luminosity
of \Lint of pp collisions collected during 2016 with the CMS detector at the CERN LHC.
Event counts in bins of the number of jets (\njets), the number of b-tagged jets (\nbtags), the scalar sum of the transverse momenta \pt of all selected jets (\Ht), and
the \mttwo variable~\cite{MT2at13TeV,MT2variable} are compared against estimates of the background from SM processes derived from dedicated data control samples.
We observe no evidence for a significant excess above the expected background event yield and interpret the results as
exclusion limits at 95\% confidence level
on the production of
pairs of gluinos and squarks using simplified models of supersymmetry (SUSY)~\cite{Ramond:1971gb,Golfand:1971iw,Neveu:1971rx,
Volkov:1972jx,Wess:1973kz,Wess:1974tw,Fayet:1974pd,Nilles:1983ge}.
Model-independent limits on the number of non-SM events are also provided for a simpler set of inclusive search regions.
\section{The CMS detector}

The central feature of the CMS apparatus is a superconducting solenoid of 6\unit{m} internal diameter,
providing a magnetic field of 3.8\unit{T}. Within the solenoid volume are a silicon pixel and strip tracker, a
lead tungstate crystal electromagnetic calorimeter, and a brass and scintillator hadron calorimeter, each composed of
a barrel and two endcap sections. Forward calorimeters extend the pseudorapidity ($\eta$) coverage provided by the
 barrel and endcap detectors. Muons are measured in gas-ionization detectors embedded in the steel flux-return yoke outside
the solenoid. The first level of the CMS trigger system, composed of custom hardware processors, uses information from the
calorimeters and muon detectors to select the most interesting events in a fixed time interval of less than 4\mus. The high-level
 trigger processor farm further decreases the event rate from around 100\unit{kHz} to less than 1\unit{kHz}, before data storage.
A more detailed description of the CMS detector and trigger system, together with a definition of the coordinate system used and the relevant
kinematic variables, can be found in Refs.~\cite{Chatrchyan:2008zzk,CMStrigger}.

\section{Event selection and Monte Carlo simulation}
\label{sec:evtsel}
Events are processed using the particle-flow (PF)
algorithm~\cite{pflowNEW},
which is designed to reconstruct and identify all particles using
the optimal combination of information from the elements of
the CMS detector.
Physics objects reconstructed with this algorithm are hereafter referred to as particle-flow candidates.
The physics objects and the event preselection are similar to those described in Ref.~\cite{MT2at13TeV}, and are summarized in Table~\ref{tab:evtsel}.
We select events with at least one jet, and veto events with an isolated lepton ($\Pe$ or $\mu$) or charged PF candidate.
The isolated charged PF candidate selection is designed to provide additional rejection against events with electrons and muons, as well as to reject hadronic tau decays.
Jets are formed by clustering PF candidates using the
anti-\kt algorithm~\cite{Cacciari:2008gp, Cacciari:2011ma}
and are
corrected for contributions from event pileup~\cite{cacciari-2008-659}
and the effects of non-uniform detector response.
Only jets passing the selection criteria in Table~\ref{tab:evtsel} are used for counting and the determination of kinematic variables.
Jets consistent with originating from a heavy-flavor hadron are identified using the combined secondary vertex tagging algorithm~\cite{btagRun2},
with a working point chosen such that the efficiency to identify a \PQb
quark jet is in the range 50--65\% for jet \pt
between 20 and 400\GeV.
The misidentification rate is approximately 1\% for light-flavor and gluon jets and 10\% for charm jets.
A more detailed discussion of the algorithm performance is given in Ref.~\cite{btagRun2}.

The negative of the vector sum of the \pt of all selected jets is denoted
by \vMht, while \vMet is defined as the negative of the vector \pt sum of
all reconstructed PF candidates.
The jet corrections are also used to correct \vMet.
Events with possible contributions from beam-halo processes or anomalous noise in the calorimeter
are rejected using dedicated filters~\cite{Chatrchyan:2011tn,MetFiltersRun2}.
For events with at least two jets, we start with the pair having the largest dijet invariant mass
and iteratively cluster all selected jets using a hemisphere algorithm that minimizes the Lund
distance measure~\cite{LundDistRef1,Phythia64} until two stable pseudo-jets are obtained.
The resulting pseudo-jets together with the \vMet are used to calculate the kinematic variable \mttwo as:
\begin{equation}
\mttwo = \min_{\vMet{}^{ \mathrm{X}(1)} + \vMet{}^{ \mathrm{X}(2)} = \vMet}
  \left[ \max \left( \Mt^{(1)} , \Mt^{(2)} \right) \right],
\label{eq.MT2.definition}
\end{equation}
where $\vMet{}^{ \mathrm{X}(i)}$ ($i=1$,2) are trial vectors
obtained by decomposing \vMet,  and
$\Mt^{(i)}$ are the transverse masses obtained by pairing either of the trial vectors with one of the two pseudo-jets.
The minimization is performed over all trial momenta satisfying the \vMet constraint.
The background from multijet events (discussed in Sec.~\ref{sec:bkgds}) is characterized by small values of
\mttwo, while larger \mttwo values are obtained in processes with significant, genuine \vMet.

\begin{table*}[tbhp]
  \setlength{\extrarowheight}{.7em}
  \centering
    \topcaption{\label{tab:evtsel} Summary of reconstruction objects and
      event preselection.  Here
$R$ is the distance parameter of the anti-\kt algorithm.
      For veto leptons and tracks, the transverse mass \Mt is determined using the veto object and the \vMet.
      The variable $\pt^{\text{sum}}$ is a measure of isolation and it denotes the sum of the transverse momenta of all the PF candidates in a cone around the lepton or the track.
The size of the cone, in units of $\Delta R \equiv \sqrt{\smash[b]{(\Delta \phi)^2
  + (\Delta \eta)^2}}$ is given in the table.
      Further details of the lepton selection are described in Ref.~\cite{MT2at13TeV}.
      The $i$th highest-$\pt$ jet is denoted as j$_i$.}
    \begin{tabular}{ l | l }
      \hline
      \multirow{3}{*}{Trigger} & $\Met>120\GeV$ and $\Mht>120\GeV$ or \\
      & $\Ht>300\GeV$ and $\Met>110\GeV$ or \\
      & $\Ht>900\GeV$ or jet $\pt>450\GeV$ \\  \hline
      Jet selection & $R=0.4$, $\pt>30\GeV$, $\abs{\eta}<2.4$ \\ \hline
      \PQb tag selection & $\pt>20\GeV$, $\abs{\eta}<2.4$ \\  \hline
      \multirow{3}{*}{$\Met$} & $\Met>250\GeV$ for $\Ht<1000\GeV$,
      else $\Met>30\GeV$\\
      & $\dpmin = \Delta\phi\left(\ptmiss, j_{\mathrm{1,2,3,4}}\right)>0.3$ \\
      & $\abs{\vMet-\vMht}/\Met<0.5$ \\ \hline
      \mttwo & $\mttwo>200\GeV$ for $\Ht<1500\GeV$, else
      $\mttwo>400\GeV$ \\ \hline
      \multirow{2}{*}{Veto muon} & $\pt>10\GeV$, $\abs{\eta}<2.4$, $\pt^{\text{sum}} < 0.2 \,\pt^{\text{lep}}$ or \\
      & $\pt>5\GeV$, $\abs{\eta}<2.4$, $\Mt<100\GeV$, $\pt^{\text{sum}}
      < 0.2 \, \pt^{\text{lep}}$ \\ \hline
      \multirow{2}{*}{Veto electron} & $\pt>10\GeV$, $\abs{\eta}<2.4$, $\pt^{\text{sum}} < 0.1 \, \pt^{\text{lep}}$ or \\
      & $\pt>5\GeV$, $\abs{\eta}<2.4$, $\Mt<100\GeV$, $\pt^{\text{sum}}
      < 0.2 \, \pt^{\text{lep}}$ \\ \hline
      Veto track & $\pt>10\GeV$, $\abs{\eta}<2.4$, $\Mt<100\GeV$,
      $\pt^{\text{sum}} < 0.1 \, \pt^{\text{track}}$ \\ \hline
\multirow{2}{*}{$\pt^{\text{sum}} $ cone} & Veto e or \Pgm: $\Delta R= \min(0.2, \max(10\GeV/\pt^{\text{lep}},0.05)) $ \\
    & Veto track: $\Delta R=$ 0.3 \\
      \hline
    \end{tabular}
\end{table*}

Collision events are selected using triggers with requirements on \Ht, \Met, \Mht, and jet \pt.
The combined trigger efficiency, as measured in a data sample of events with an isolated electron,
is found to be $>$98\% across the full kinematic range of the search.
To suppress background from multijet production, we require $\mttwo >$ 200\GeV in events with $\njets \geq 2$ and $\Ht < 1500$\GeV.
This \mttwo threshold is increased to 400\GeV for events with $\Ht > 1500$\GeV to maintain
multijet processes as a subdominant background in all search regions.
To protect against jet mismeasurement, we require the minimum difference in azimuthal angle between the \vMet  vector and
each of the leading four jets, \dpmin, to be greater than 0.3, and the magnitude of the difference between \vMet  and \vMht  to be less than half of \Met.
For the determination of \dpmin we consider jets with $\abs{\eta}<4.7$.
If less than four such jets are found, all are considered in the \dpmin calculation.

Events containing at least two jets are categorized by the values of \njets, \nbtags, and \Ht.
Each such bin is referred to as a \emph{topological region}.
Signal regions are defined by further dividing topological regions into bins of \mttwo.
Events with only one jet are selected if the \pt of the jet is at least 250\GeV,
and are classified according to the \pt of this jet and whether the event contains a b-tagged jet.
The search regions are summarized in Tables~\ref{tab:sr1}-\ref{tab:sr3} in Appendix~\ref{app:srs}.
We also define \emph{super signal regions}, covering a subset of the kinematic space of the full analysis with simpler inclusive selections.
The super signal regions can be used to obtain approximate interpretations of our
result, as discussed in Section~\ref{sec:results}, where these
regions are defined.

{\tolerance=1200
Monte Carlo (MC) simulations are used to design the search,
to aid in the estimation of SM backgrounds, and to evaluate the sensitivity
to gluino and squark pair production in simplified models of SUSY.
The main background samples (\zjets, \wjets, and \ttjets),
as well as signal samples of gluino and squark pair production,
are generated at leading order (LO) precision with the \MADGRAPH~5 generator~\cite{mg5amcnlo,mlm} interfaced with \PYTHIA 8.2~\cite{pythia8}
for fragmentation and parton showering.
Up to four, three, or two additional partons are considered in the matrix element calculations for the generation of the \vjets,
\ttjets, and signal samples, respectively.
Other background processes are also considered: \ttV samples are generated at LO precision with the \MADGRAPH~5 generator, with up to two additional partons in the matrix element calculations,
while single top samples are generated at next-to-leading order (NLO) precision with the \MADGRAPH{\textunderscore}aMC@NLO~\cite{mg5amcnlo} or \POWHEG~\cite{Alioli:2009je,Re:2010bp} generators.
Contributions from rarer processes such as diboson, triboson, and four top production, are found to be negligible.
Standard model samples are simulated with a detailed \GEANTfour~\cite{geant4} based detector simulation and processed using the same chain of reconstruction programs as collision data,
while the CMS fast simulation program~\cite{fastsim} is used for the signal samples.
The most precise available cross section calculations are used to normalize the simulated samples, corresponding most often to
NLO or next-to-NLO accuracy~\cite{Gavin:2010az, Gavin:2012sy, Czakon:2011xx, mg5amcnlo, Alioli:2009je, Re:2010bp, Borschensky:2014cia}.
\par}

To improve on the \MADGRAPH modeling of the
multiplicity of additional jets from initial state radiation (ISR),
\MADGRAPH \ttbar MC events are weighted based on the
number of ISR jets ($N_\mathrm{j}^\mathrm{ISR}$) so as to make the jet
multiplicity agree with data.
The same reweighting procedure is applied to SUSY MC events.
The weighting factors are obtained from a control region enriched in \ttbar,
obtained by selecting events with two leptons and exactly two b-tagged jets, and vary between 0.92 for $N_\mathrm{j}^\mathrm{ISR}=1$ and 0.51 for $N_\mathrm{j}^\mathrm{ISR}\geq 6$.
We take one half of the deviation from unity as the systematic uncertainty in these reweighting factors, to cover for differences between \ttbar and SUSY production.
\section{Backgrounds}
\label{sec:bkgds}

The backgrounds in jets-plus-\Met final states typically arise from three categories of SM processes:

\begin{itemize}
\item ``lost lepton (LL)'', i.e., events with a lepton from a \PW\ decay where the lepton is either out of acceptance, not reconstructed, not identified, or not isolated.

  This background originates mostly from  \wjets and \ttjets events, with smaller contributions from rarer processes such as diboson or \ttV production.
\item ``irreducible'', i.e., \zjets events, where the \PZ\ boson decays to neutrinos.  This background is most similar to potential signals.
  It is a major background in nearly all search regions, its
  importance decreasing with
increasing \nbtags.
\item ``instrumental background'', i.e., mostly multijet events with no genuine \Met.
  These events enter a search region due to either significant jet momentum mismeasurements, or sources of anomalous noise.
\end{itemize}

\subsection{Estimation of the background from events with leptonic \PW\ boson decays}
\label{sec:bkgds:ll}

Control regions with exactly one lepton candidate are selected using
the same triggers
and preselections used for the signal regions,
with the exception of the lepton veto, which is inverted.
Selected events are binned according to the same criteria as the search regions,
and the background in each signal bin,
$N^{\mathrm{SR}}_{\mathrm{LL}}$, is obtained
from the number of events in the control region,
$N^{\mathrm{CR}}_{1\ell}$,
using transfer factors according to:
\ifthenelse{\boolean{cms@external}}{
\begin{multline}
\label{eq:ll}
  N^{\mathrm{SR}}_{\mathrm{LL}} \left(\Ht,\njets,\nbtags,\mttwo\right) = \\
  N^{\mathrm{CR}}_{1\ell} \left(\Ht,\njets,\nbtags,\mttwo\right) \\
  \times R^{0\ell/1\ell}_{\mathrm{MC}} \left(\Ht,\njets,\nbtags,\mttwo\right) \, k \left(\mttwo\right).
\end{multline}
}{
\begin{equation}
\label{eq:ll}
  N^{\mathrm{SR}}_{\mathrm{LL}} \left(\Ht,\njets,\nbtags,\mttwo\right) = N^{\mathrm{CR}}_{1\ell} \left(\Ht,\njets,\nbtags,\mttwo\right) \, R^{0\ell/1\ell}_{\mathrm{MC}} \left(\Ht,\njets,\nbtags,\mttwo\right) \, k \left(\mttwo\right).
\end{equation}
}
The single-lepton control region typically has 1--2 times as many events as the corresponding signal region.
The factor $R^{0\ell/1\ell}_{\mathrm{MC}} \left(\Ht,\njets,\nbtags,\mttwo\right)$ accounts for lepton acceptance and efficiency
and the expected contribution from the decay of \PW\ bosons to hadrons through an intermediate \Pgt\ lepton.
It is obtained from MC simulation, and corrected for measured
differences in lepton efficiencies between data and simulation.

\begin{figure}[htb]
  \centering
    \includegraphics[width=0.48\textwidth]{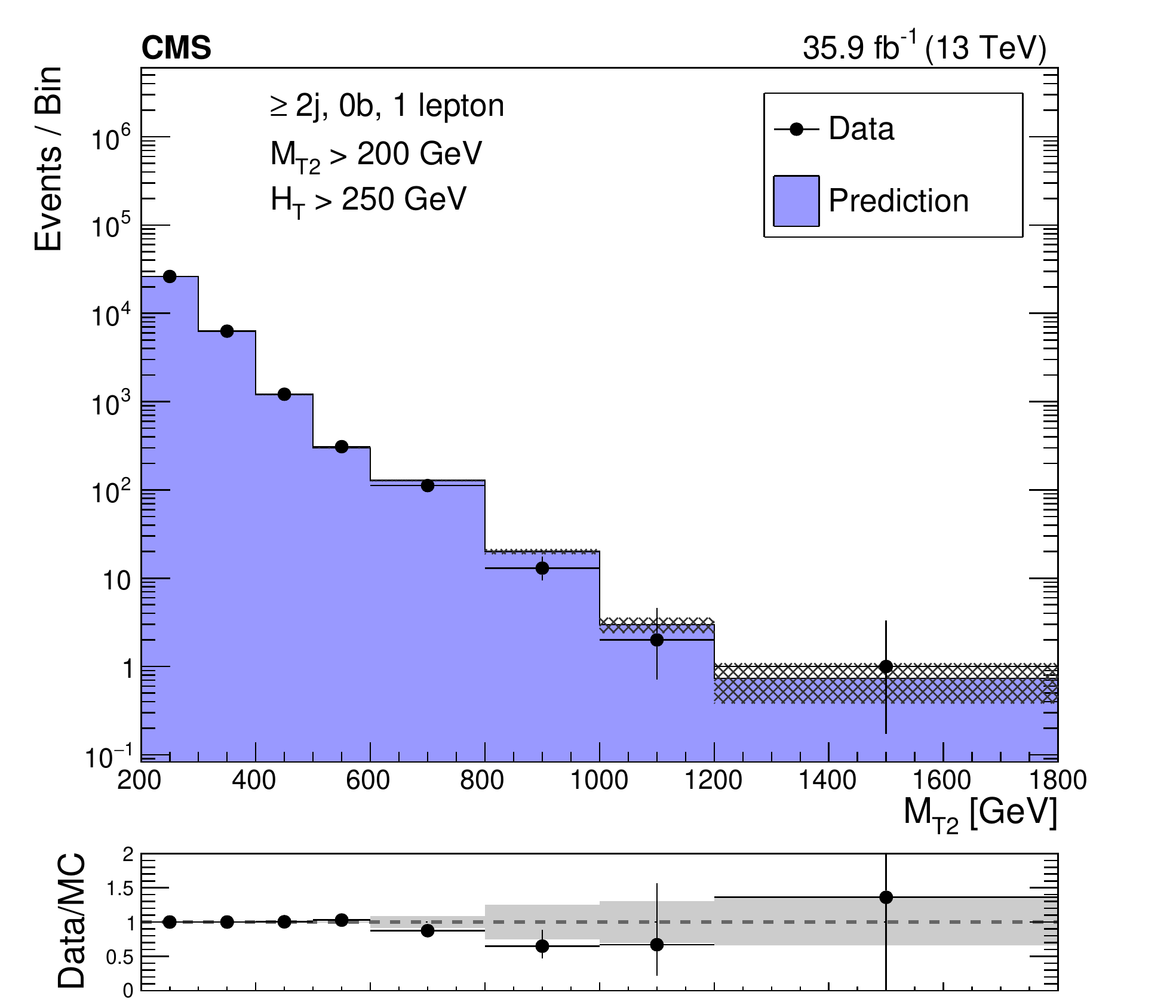}
    \includegraphics[width=0.48\textwidth]{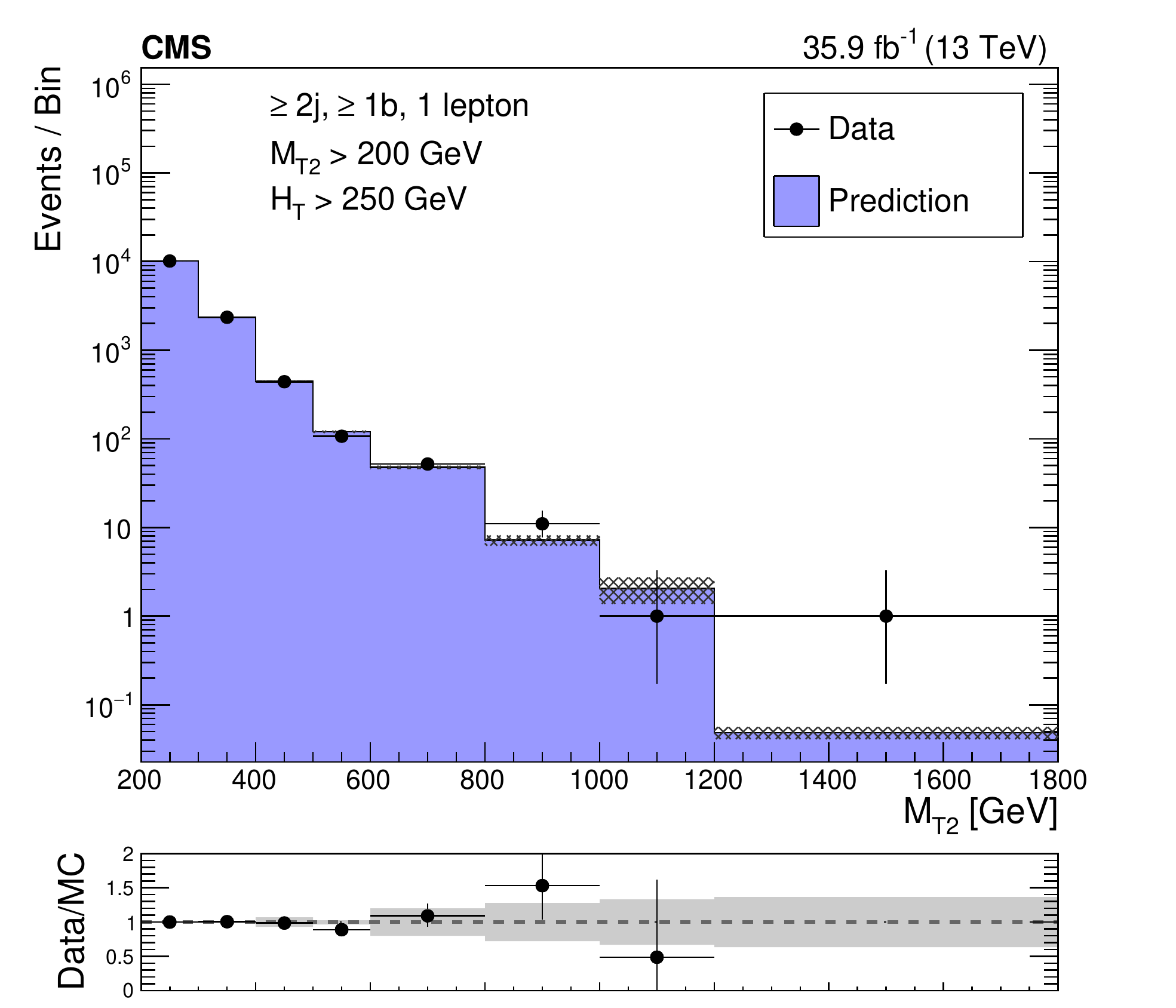}
    \caption{Distributions of the \mttwo variable in data and simulation for the single-lepton control region selection, after
      normalizing the simulation to data in the control region bins of \Ht, \njets, and \nbtags
      for events with no \PQb-tagged jets (\cmsLeft), and events with at
      least one \PQb-tagged jet (\cmsRight).
      The hatched bands on the top panels show the MC statistical uncertainty,
      while the solid gray bands in the ratio plots show the systematic uncertainty in the \mttwo shape.}
    \label{fig:ll_mt2}
\end{figure}

The factor $k\left(\mttwo\right)$ accounts for the distribution, in bins of \mttwo, of the estimated background in each topological region.
It is obtained using both data and simulation as follows.
In each topological region, the control region corresponding to the highest \mttwo bin is successively combined with the next highest bin
until the expected SM yield in combined bins is at least 50 events.
When two or more control region bins are combined, the fraction of events expected to populate a particular \mttwo bin,  $k \left(\mttwo\right)$,
is determined using the expectation from SM simulated samples, including all relevant processes.
The modeling of \mttwo is checked in data using single-lepton control samples enriched in events originating from either \wjets or \ttjets,
as shown in the \cmsLeft and \cmsRight panels of Fig.~\ref{fig:ll_mt2}, respectively.
The predicted distributions in the comparison are obtained by summing all control regions after normalizing MC yields to data and distributing events among
\mttwo bins according to the expectation from simulation, as is done for the estimate of the lost-lepton background.
For events with $\njets =$ 1, a control region is defined for each bin of jet \pt.

Uncertainties from the limited size of the control sample and
from theoretical and experimental sources
are evaluated and propagated to the final estimate.
The dominant uncertainty in $R^{0\ell/1\ell}_{\mathrm{MC}} \left(\Ht,\njets,\nbtags,\mttwo\right)$ arises from the modeling
of the lepton efficiency (for electrons, muons, and hadronically-decaying tau leptons) and jet energy scale (JES) and is of order 15--20\%.
The uncertainty in the \mttwo extrapolation, which is as large as 40\%, arises primarily from the JES,
the relative fractions of \wjets\ and \ttjets,
and variations of the renormalization and factorization scales assumed for their simulation.
These and other uncertainties are similar to those in Ref.~\cite{MT2at13TeV}.

\subsection{Estimation of the background from \texorpdfstring{$\PZ(\PGn\PAGn)+$jets}{PZ nu nu-bar + jets}}
\label{sec:bkgds:zinv}

The $\PZ\to\PGn\PAGn$ background is estimated from a dilepton control sample selected using triggers requiring two leptons.
The trigger efficiency, measured with a data sample of events with large \HT, is found to be greater than 97\% in the selected kinematic range.
To obtain a control sample enriched in $\PZ\to\ell^{+}\ell^{-}$ events ($\ell = \Pe,\mu$),
we require that the leptons are of the same flavor, opposite charge, that the \pt of the leading and trailing leptons are at least 100~GeV and 30~GeV, respectively,
and that the invariant mass of the lepton pair is consistent with the mass of a \PZ\ boson within 20~GeV.
After requiring that the \pt of the dilepton system is at least 200\GeV, the preselection requirements are applied based on kinematic variables
recalculated after removing the dilepton system from the event to replicate the $\PZ\to\PGn\PAGn$ kinematics.
For events with $\njets = 1$, one control region is defined for each bin of jet \pt.
For events with at least two jets, the selected events are binned in \Ht, \njets, and \nbtags,
but not in \mttwo, to increase the dilepton event yield in each control region.

The contribution to each control region from flavor-symmetric processes, most importantly \ttbar, is estimated
using opposite-flavor (OF) $\Pe\Pgm$ events obtained with the same selections
as same-flavor (SF) $\Pe\Pe$ and  $\Pgm\Pgm$ events.
The background in each signal bin is then obtained using transfer factors according to:
\ifthenelse{\boolean{cms@external}}{
\begin{multline}
  \label{eq:zinv}
N^{\mathrm{SR}}_{\PZ\to\PGn\PAGn} \left(\Ht,\njets,\nbtags,\mttwo\right) = \Bigl[N^{\mathrm{CRSF}}_{\ell\ell} \left(\Ht,\njets,\nbtags\right)\\
 - N^{\mathrm{CROF}}_{\ell\ell} \left(\Ht,\njets,\nbtags\right) \, R^{\mathrm{SF}/\mathrm{OF}} \Bigr] \\
\times R^{\PZ\to\PGn\PAGn/Z\to\ell^{+}\ell^{-}}_{\mathrm{MC}} \left(\Ht,\njets,\nbtags\right) \, k\left(\mttwo\right).
\end{multline}
}{
\begin{multline}
  \label{eq:zinv}
N^{\mathrm{SR}}_{\PZ\to\PGn\PAGn} \left(\Ht,\njets,\nbtags,\mttwo\right) = \Bigl[N^{\mathrm{CRSF}}_{\ell\ell} \left(\Ht,\njets,\nbtags\right)
 - N^{\mathrm{CROF}}_{\ell\ell} \left(\Ht,\njets,\nbtags\right) \, R^{\mathrm{SF}/\mathrm{OF}} \Bigr] \\
\times R^{\PZ\to\PGn\PAGn/Z\to\ell^{+}\ell^{-}}_{\mathrm{MC}} \left(\Ht,\njets,\nbtags\right) \, k\left(\mttwo\right).
\end{multline}
}

\begin{figure}[htb]
  \centering
    \includegraphics[width=0.48\textwidth]{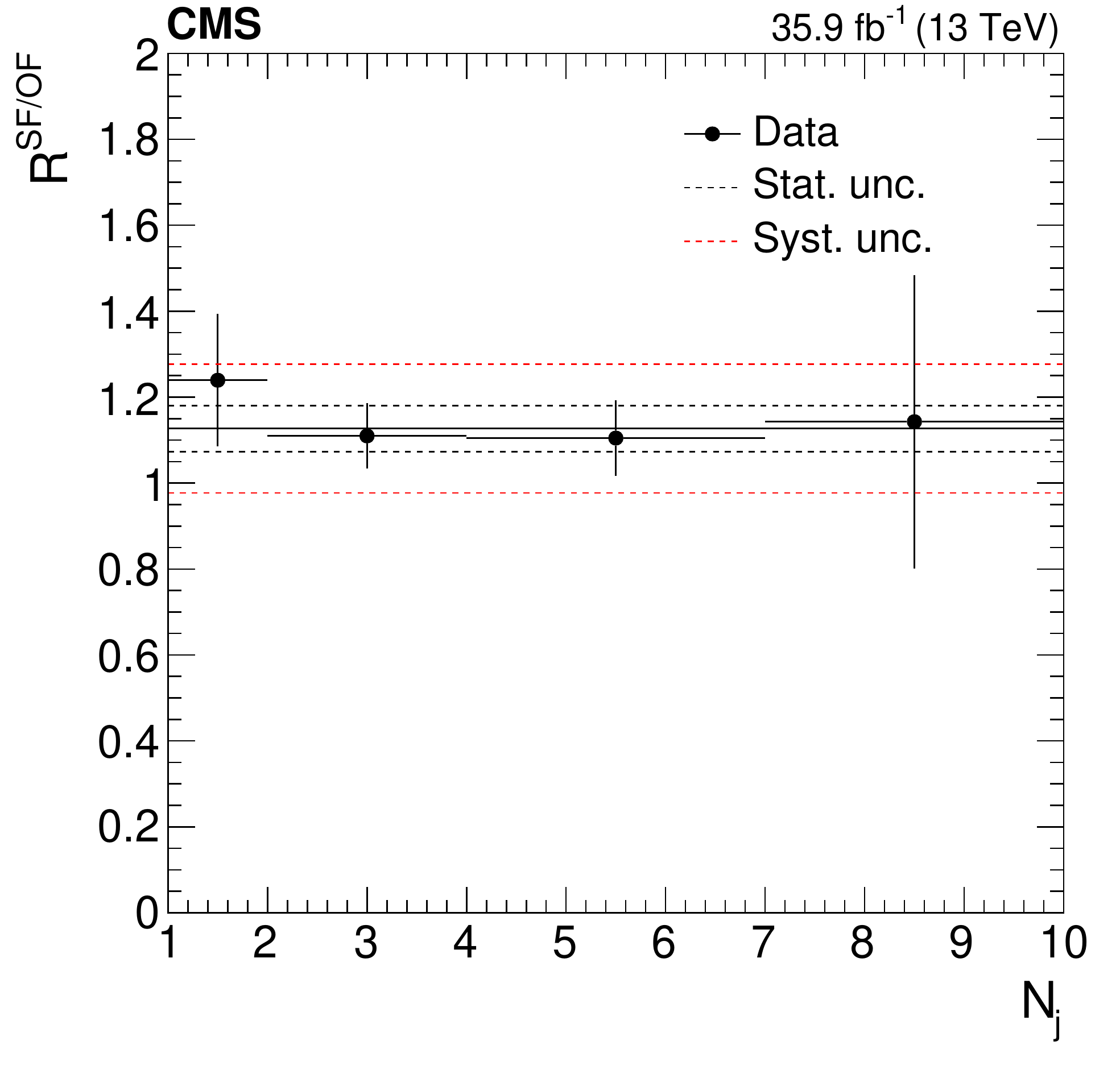}
    \includegraphics[width=0.48\textwidth]{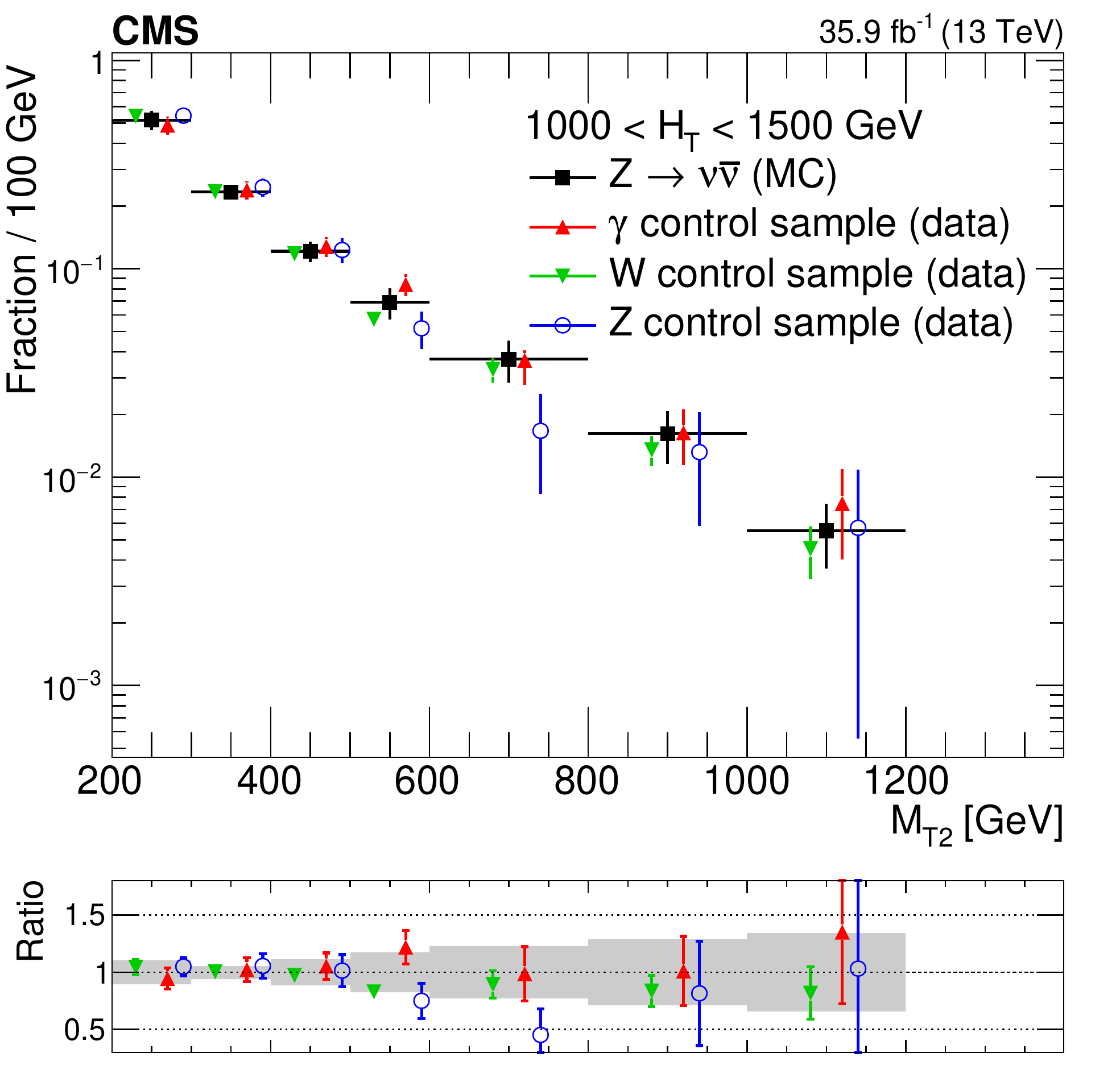}\\
    \caption{(\cmsLLeft) Ratio $R^{\mathrm{SF}/\mathrm{OF}}$ in data as a function of \njets.
      The solid black line enclosed by the red dashed lines corresponds to a value of $1.13\pm0.15$ that is observed to be stable with respect to event kinematics,
      while the two dashed black lines denote the statistical uncertainty in the $R^{\mathrm{SF}/\mathrm{OF}}$ value.
      (\cmsRRight) The shape of the \mttwo distribution in \znunu simulation compared to shapes from \cPgg, \PW, and \PZ\ data control samples in a
      region with $1000<\Ht<1500\GeV$ and $\njets\ge2$,  inclusive in \nbtags. The solid gray band on the ratio plot shows the systematic uncertainty in the \mttwo shape.}
    \label{fig:zinv}
\end{figure}

\begin{sloppypar}
Here $N^{\mathrm{CRSF}}_{\ell\ell}$ and $N^{\mathrm{CROF}}_{\ell\ell}$ are the number of
SF and OF events in the control region, while
$R^{\PZ\to\PGn\PAGn/\PZ\to\ell^{+}\ell^{-}}_{\mathrm{MC}}$ and
$k\left(\mttwo\right)$ are defined below.
The factor $R^{\mathrm{SF}/\mathrm{OF}}$ accounts for the difference in acceptance and efficiency
between SF and OF events.
It is determined as the ratio of the number of SF events to OF events in a \ttbar enriched control sample,
obtained with the same selections as the $\PZ\to\ell^{+}\ell^{-}$ sample,
but inverting the requirements on the \pt and the invariant mass of the lepton pair.
A measured value of $R^{\mathrm{SF}/\mathrm{OF}}=1.13\pm0.15$ is observed to be stable with respect to event kinematics, and is applied in all regions.
Figure~\ref{fig:zinv} (left) shows $R^{\mathrm{SF}/\mathrm{OF}}$ measured as a function of the number of jets.
\end{sloppypar}

An estimate of the $\PZ\to\PGn\PAGn$ background in each topological region is obtained from the corresponding dilepton control region
via the factor $R^{\PZ\to\PGn\PAGn/\PZ\to\ell^{+}\ell^{-}}_{\mathrm{MC}}$, which accounts for the acceptance and efficiency to select the dilepton pair
and the ratio of branching fractions for $\PZ\to\ell^{+}\ell^{-}$ and $\PZ\to\PGn\PAGn$ decays.
This factor
is obtained from simulation,
including corrections for differences in the lepton efficiencies between data and simulation.

The factor $k\left(\mttwo\right)$ accounts for the distribution, in bins of \mttwo, of the estimated background in each topological region.
This distribution is constructed using the \mttwo shape from dilepton data and \znunu simulation in each topological region.
Studies with simulated samples indicate that the \mttwo shape for \znunu events is independent of \nbtags for a given \Ht and \njets selection, and that
the shape is also independent of the number of jets for $\Ht>1500\GeV$.
The MC modeling of \nbtags and \njets as well as of the \mttwo shape in bins of \njets and \nbtags is validated in data, using a dilepton control sample.
As a result, \mttwo templates for topological regions differing only in \nbtags are combined, separately for data and simulation.
For $\Ht>1500\GeV$, only one \mttwo template is constructed for data and one for simulation by combining all relevant topological regions.

Starting from the highest \mttwo bin in each control region, we merge bins until the sum of the merged bins contains at least 50 expected events
from simulation.
The fraction of events in each uncombined bin is determined using the corresponding \mttwo template from dilepton data, corrected
by the ratio $R^{Z\to\PGn\PAGn/Z\to\ell^{+}\ell^{-}}_{\mathrm{MC}}$.
The \mttwo shape from simulation is used to distribute events among the combined bins, after normalizing the simulation to the data yield in the same group of bins.

The modeling of \mttwo is validated in data using control samples enriched in $\cPgg$, $\PW\to\ell\nu$, and $\PZ\to\ell^{+}\ell^{-}$ events in each bin of \Ht.
The \cmsRight panel of Fig.~\ref{fig:zinv} shows agreement between the \mttwo distributions obtained from $\cPgg$, $\PW$, and $\PZ$
data control samples with that from \znunu simulation for events with $1000<\Ht<1500\GeV$.
In this comparison, the \cPgg\ sample is obtained by selecting events with $\pt^{\gamma}>180$\GeV and is corrected for contributions from multijet events and $R^{\PZ/\cPgg}_{\mathrm{MC}}$,
the \PW\ sample is corrected for $R^{\PZ/\PW}_{\mathrm{MC}}$,
both the \PW\ and \PZ\ samples are corrected for contributions from top quark events,
and the \PZ sample is further corrected for $R^{\PZ\to\PGn\PAGn/\PZ\to\ell^{+}\ell^{-}}_{\mathrm{MC}}$.
Here $R^{\PZ/\cPgg}_{\mathrm{MC}}$ ($R^{\PZ/\PW}_{\mathrm{MC}}$) is the ratio
of the \mttwo distributions for \PZ boson and \cPgg\ (\PW) boson events derived in
simulation.

The largest uncertainty in the estimate of the invisible \PZ\ background in most regions results from the limited size of the dilepton control sample.
This uncertainty, as well as all other relevant theoretical and experimental uncertainties, are evaluated and propagated to the final estimate.
The dominant uncertainty in the ratio $R^{Z\to\PGn\PAGn/Z\to\ell^{+}\ell^{-}}_{\mathrm{MC}}$
is obtained from measured differences in lepton efficiency between data and simulation, and is about 5\%.
The uncertainty in the $k\left(\mttwo\right)$ factor arises from data statistics for uncombined bins,
while for combined bins it is due to uncertainties in the JES and variations in the renormalization and factorization scales.  These can result in effects as large as 40\%.
\subsection{Estimation of the multijet background}
\label{sec:bkgds:qcd}

For events with at least two jets, a multijet-enriched control region is obtained in each \Ht bin by inverting
the \dpmin requirement described in Section~\ref{sec:evtsel}.
Events are selected using \HT triggers, and the extrapolation from low- to high-\dpmin is based on the following ratio:
\begin{equation}
\label{eq:qcd_ratio}
r_{\phi}(\mttwo) = N(\Delta\phi_{\min} > 0.3) / N(\Delta\phi_{\min} < 0.3).
\end{equation}

\begin{figure*}[tbhp]
  \centering
    \includegraphics[width=0.32\linewidth]{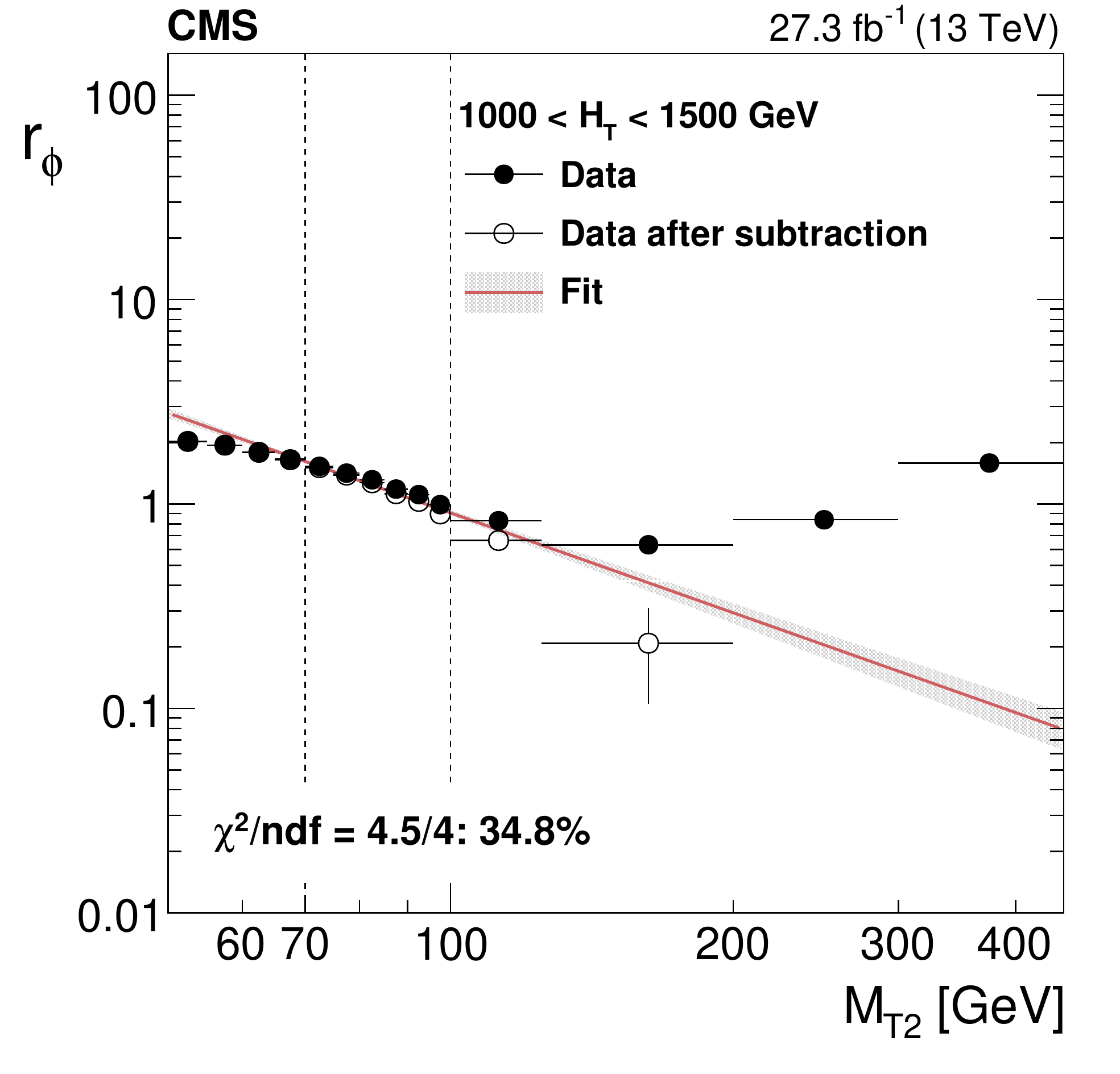}
    \includegraphics[width=0.32\linewidth]{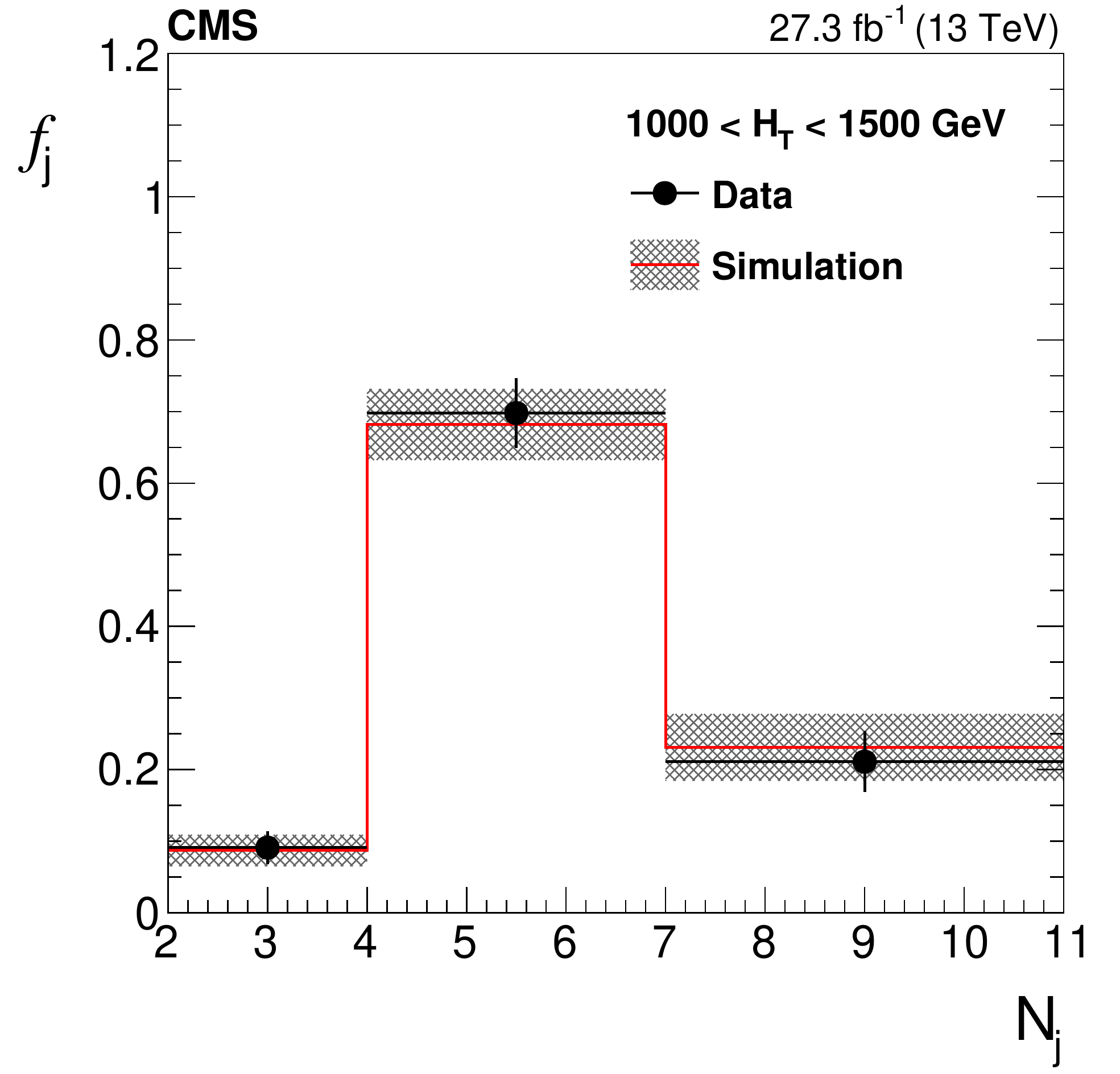}
    \includegraphics[width=0.32\linewidth]{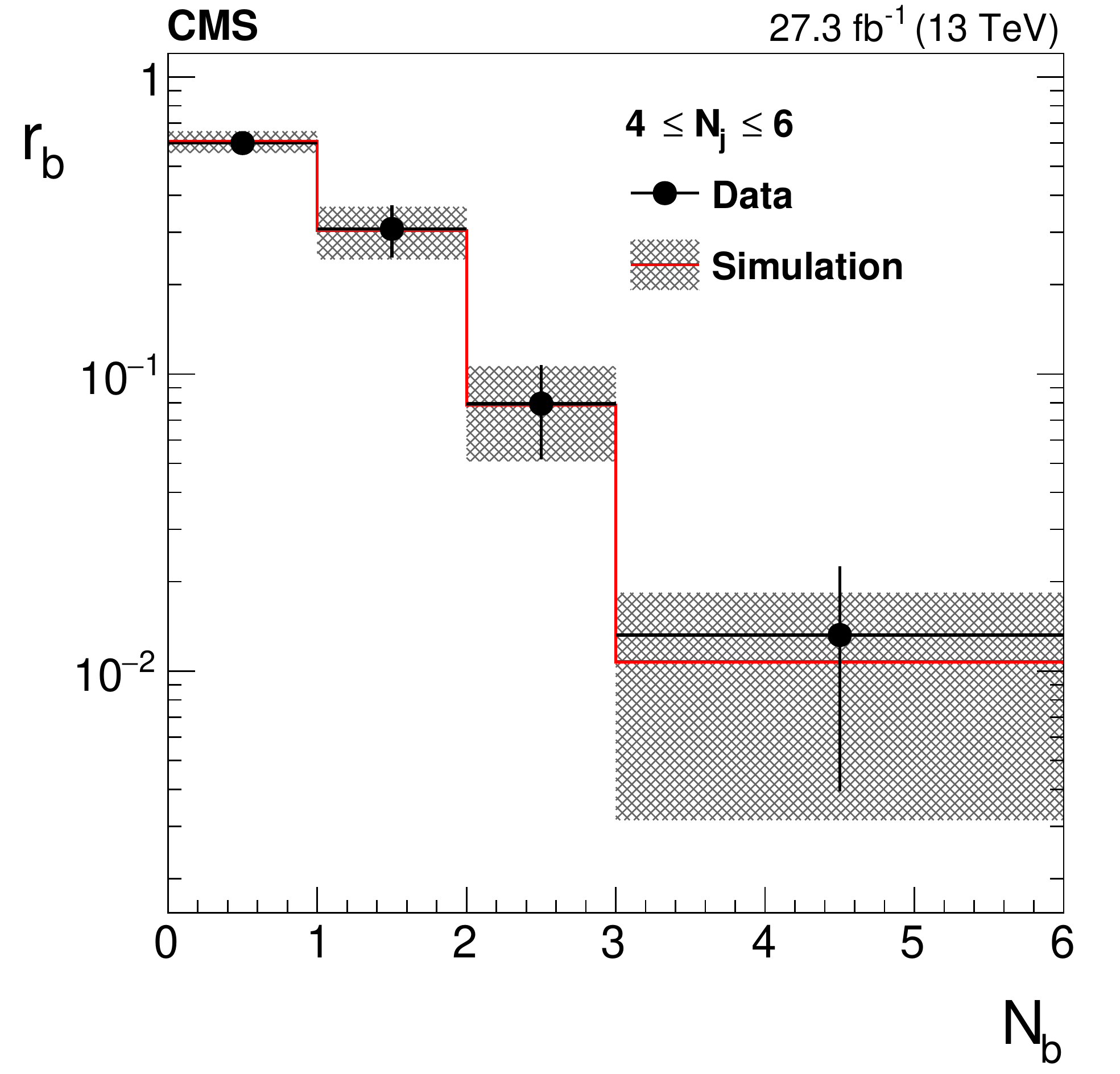}
    \caption{The ratio $r_{\phi}$ as a function of \mttwo for $1000 < \Ht < 1500$\GeV (\cmsLeft).
      The superimposed fit is performed to the open circle data points.
      The black points represent the data before subtracting non-multijet contributions using simulation.
      Data point uncertainties are statistical only.
      The red line and the grey band around it show the result of the fit to a power-law function performed in the window $70 < \mttwo <100$\GeV and the associated fit uncertainty.
      Values of $f_\mathrm{j}$, the fraction of events in bin \njets, (middle) and $r_{\PQb}$, the fraction of events
in bin \njets that fall in bin \nbtags, (\cmsRight)
      are measured in data after requiring $\dpmin < 0.3$ and $100 < \mttwo < 200$\GeV.
      The hatched bands represent both statistical and systematic uncertainties.}
    \label{fig:qcd}
\end{figure*}

Studies with simulated samples show that the ratio can be described by a power law as
$r_{\phi}(\mttwo) = a \, \mttwo^{b}$. The parameters
$a$ and  $b$
are determined separately in each \Ht bin by fitting $r_{\phi}$ in an \mttwo sideband in data after subtracting non-multijet contributions using simulation.
The sideband spans \mttwo values of 60--100\GeV for events with $\HT <$ 1000\GeV, and 70--100\GeV for events with larger values of \HT.
The fit to the $r_{\phi}$ distribution in the 1000 $< \Ht <$ 1500\GeV region is shown in Fig.~\ref{fig:qcd} (left).
The inclusive multijet contribution in each signal region, $N^\mathrm{SR}_{\mathrm{j,b}}\left(\mttwo\right)$, is estimated using the ratio $r_{\phi}(\mttwo)$ measured in the \mttwo sideband and the number of events in the low-\dpmin control region, $N^\mathrm{CR}_{\mathrm{inc}}\left(\mttwo\right)$, according to
\begin{equation}
\label{eq:qcd_inc}
  N^{\mathrm{SR}}_{\mathrm{j,b}}(\mttwo) = N^{\mathrm{CR}}_{\mathrm{inc}}\left(\mttwo\right) \, r_{\phi}(\mttwo) \, f_{\mathrm{j}} \left(\Ht\right) \, r_{\mathrm{b}} \left(\njets\right),
\end{equation}
where $f_\mathrm{j}$ is the fraction of multijet events in bin \njets, and $r_{\PQb}$ is the fraction of events in bin \njets that are in bin \nbtags. (Here, \njets denotes a jet multiplicity bin, and \nbtags denotes a \PQb jet multiplicity bin within \njets).
The values of $f_\mathrm{j}$ and $r_{\PQb}$ are measured using events with \mttwo  between 100 and 200\GeV in the low \dpmin sideband,
where $f_\mathrm{j}$ is measured separately in each \Ht bin, while $r_{\PQb}$ is measured in bins of \njets
integrated over \Ht, as $r_{\PQb}$ is found to be independent of the latter.
Values of $f_\mathrm{j}$ and $r_{\PQb}$ measured in data are shown in Fig.~\ref{fig:qcd} (center and right) compared to simulation.

The largest uncertainties in the estimate in most regions result from the statistical uncertainty in the fit
and from the sensitivity of the $r_{\phi}$ value to variations in the fit window.
These variations result in an uncertainty that increases with \mttwo and ranges from 20--50\%.
The total uncertainty in the estimate is found to be of similar size as in Ref.~\cite{MT2at13TeV},
varying between 40--180\% depending on the search region.

An estimate based on $r_{\phi}(\mttwo)$ is not viable in the monojet search regions, which therefore require a different strategy.
A control region is obtained by selecting events with a second jet with $30 <\pt < 60$\GeV and inverting the \dpmin requirement.
After subtracting non-multijet contributions using simulation, the data yield in the control region is taken as an estimate of the background in the corresponding monojet search region.
Tests in simulation show the method provides a conservative estimate of the multijet background, which is less than 8\% in all monojet search regions.
In all monojet bins, a 50\% uncertainty in the non-multijet subtraction is combined with the statistical uncertainty from the data yield in the control region with a second jet.
\section{Results}
\label{sec:results}

\begin{figure*}[!htb]
  \centering
    \includegraphics[width=\textwidth]{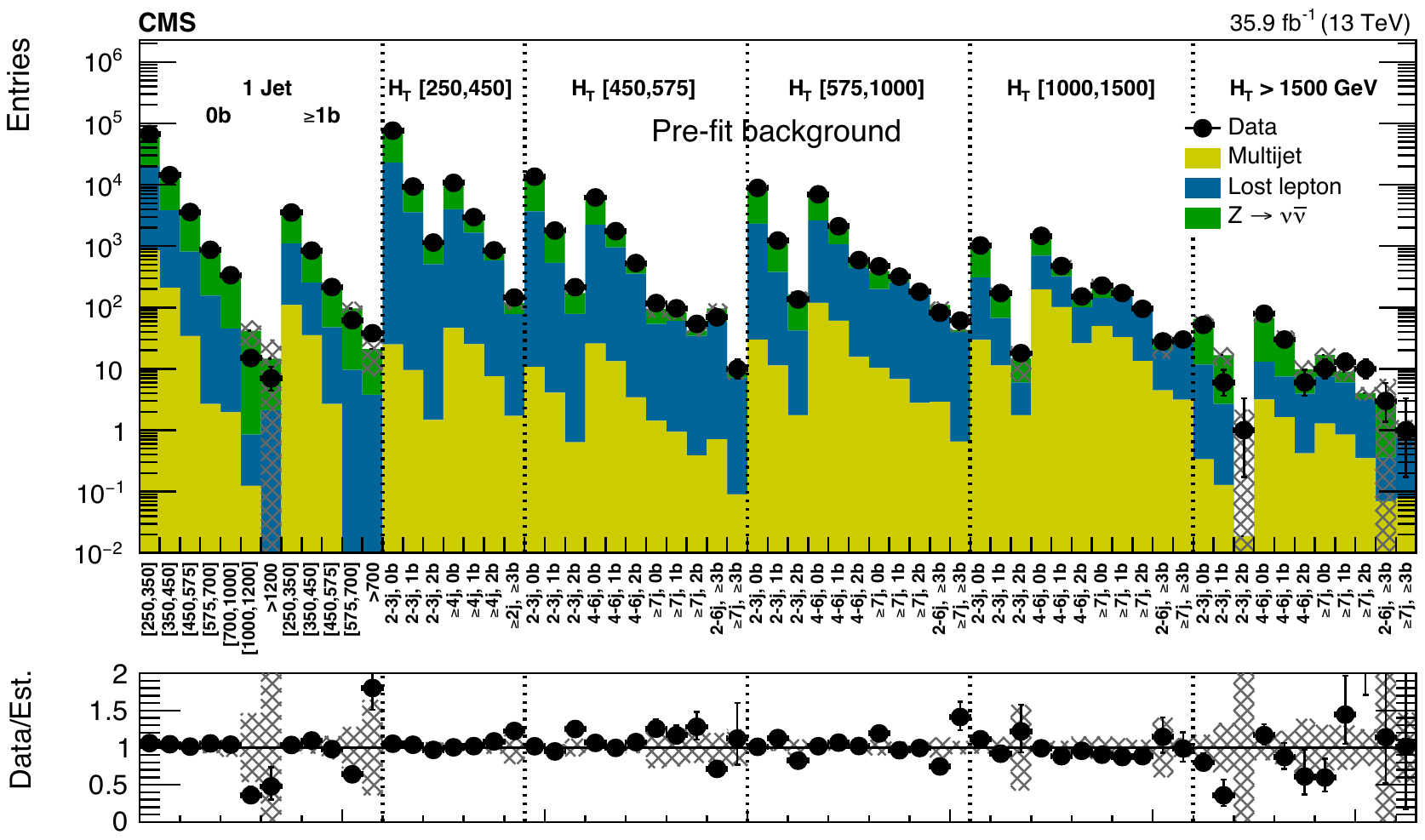}\\
    \includegraphics[width=\textwidth]{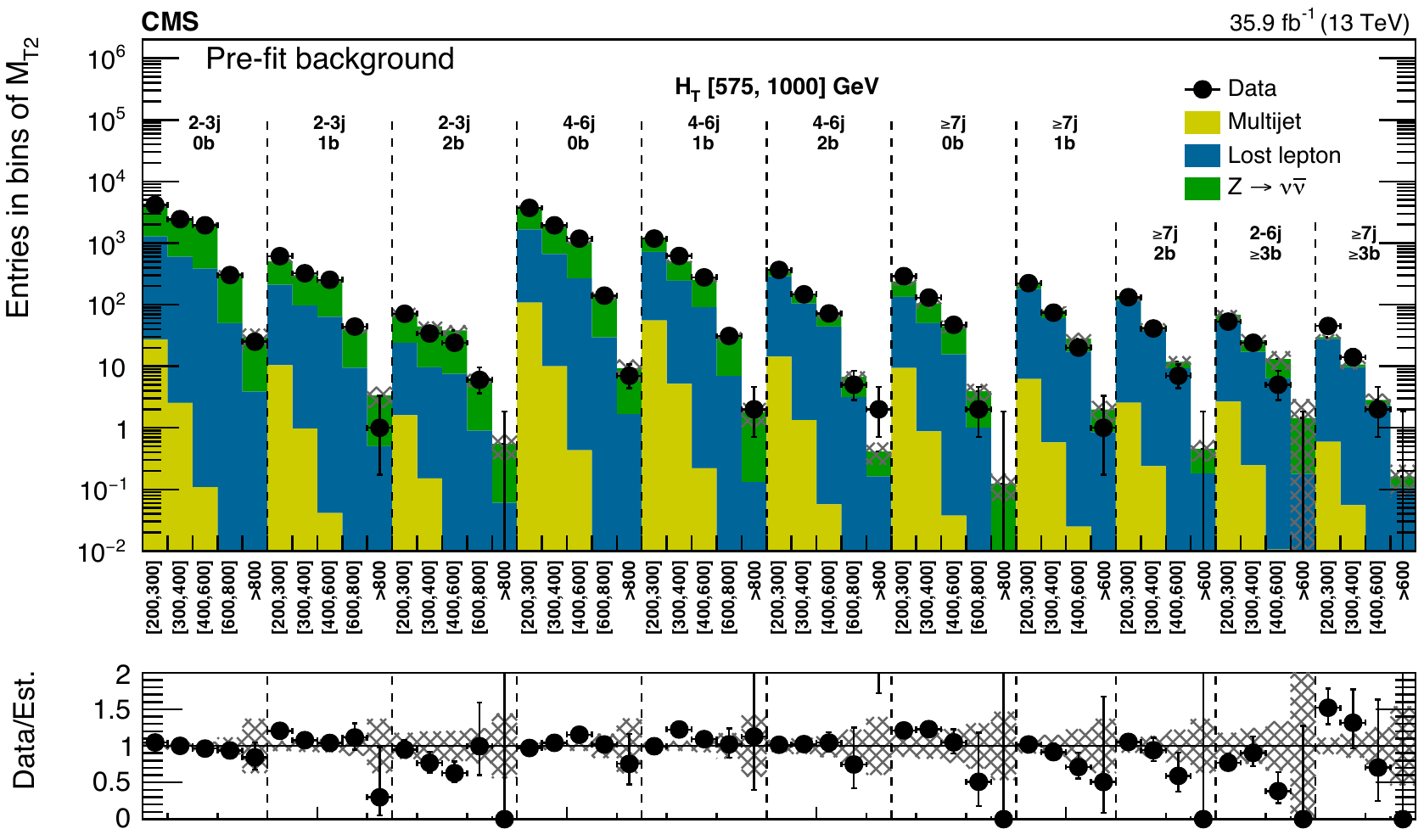}\\
    \caption{
    (\cmsLLeft) Comparison of estimated (pre-fit) background and observed data events in each topological region.  Hatched bands represent the full uncertainty in the background estimate.
      The results shown for $\njets = 1$ correspond to the monojet search regions binned in jet \pt,
      whereas for the multijet signal regions, the notations j, b indicate \njets, \nbtags labeling.
      (\cmsRRight) Same for individual \mttwo signal bins in the medium
      \Ht region. On the $x$-axis, the \mttwo binning is shown in
      units of \GeV.}
    \label{fig:results}
\end{figure*}

The data yields in the search regions are statistically compatible with the estimated
backgrounds from SM processes.
A summary of the results of this search is shown in Fig.~\ref{fig:results}.
Each bin in the \cmsLeft panel corresponds to a single \Ht, \njets,
\nbtags topological region, integrated over \mttwo.
The \cmsRight panel further breaks down the background estimates and observed data yields into \mttwo bins for the
region $575 < \Ht <1000$\GeV.
Distributions for the other \Ht regions can be found in Appendix~\ref{app:results}.
The background estimates and corresponding uncertainties shown in these plots rely exclusively
on the inputs from control samples and simulation described in
Section~\ref{sec:bkgds}, and are referred to
in the rest of the text as ``pre-fit background'' results.

To allow simpler reinterpretation, we also provide results for super signal regions, which cover subsets of the full
analysis with simpler inclusive selections and that can be used to obtain approximate interpretations of this search.
The definitions of these regions are given in Table~\ref{tab:ssr_def_yields},
with the predicted and observed number of events
and the 95\% confidence level (CL) upper limit on the number of signal events contributing to each region.
Limits are set using a modified frequentist approach, employing the CL$_{\mathrm{s}}$ criterion and relying on asymptotic approximations
to calculate the distribution of the profile likelihood test-statistic
used~\cite{Read:2002hq,Junk:1999kv,Cowan:2010js,ATLAS:2011tau}.

\begin{table*}[!htb]
  \centering
    \topcaption{\label{tab:ssr_def_yields} Definitions of super signal
      regions, along with predictions, observed data, and the observed
      95\% CL upper limits on the number of signal events contributing to
      each region ($N_{95}^{\text{obs}}$).  The limits are shown as a range
      corresponding to an assumed uncertainty in the signal acceptance of 0-15\%.
      A dash in the selections means that no requirement is applied.}
      \begin{tabular}{l|cccc ccc}
      \hline
      Region & \njets & \nbtags & \HT [\GeVns{}] & \mttwo [\GeVns{}] & Prediction & Data & $N_{95}^{\text{obs}}$ \\
      \hline
      2j loose   & $\geq$2 &\NA     & $>$1000 & $>$1200 & $38.9 \pm 11.2$ & 42 & 26.6--27.8 \\
      2j tight   & $\geq$2 &\NA     & $>$1500 & $>$1400 & $2.9 \pm 1.3$ & 4 & 6.5--6.7 \\
      \hline
      4j loose   & $\geq$4 &\NA     & $>$1000 & $>$1000 & $19.4 \pm 5.8$ & 21 & 15.8--16.4 \\
      4j tight   & $\geq$4 &\NA     & $>$1500 & $>$1400 & $2.1 \pm 0.9$ & 2 & 4.4--4.6 \\
      \hline
      7j loose   & $\geq$7 &\NA     & $>$1000 & $>$600  & $23.5^{+5.9}_{-5.6}$ & 27 & 18.0--18.7 \\
      7j tight   & $\geq$7 &\NA     & $>$1500 & $>$800  & $3.1^{+1.7}_{-1.4}$ & 5 & 7.6--7.9 \\
      \hline
      2b loose   & $\geq$2 & $\geq$2 & $>$1000 & $>$600  & $12.9^{+2.9}_{-2.6}$ & 16 & 12.5--13.0 \\
      2b tight   & $\geq$2 & $\geq$2 & $>$1500 & $>$600  & $5.1^{+2.7}_{-2.1}$ & 4 & 5.8--6.0 \\
      \hline
      3b loose   & $\geq$2 & $\geq$3 & $>$1000 & $>$400  & $8.4 \pm 1.8$ & 10 & 9.3--9.7 \\
      3b tight   & $\geq$2 & $\geq$3 & $>$1500 & $>$400  & $2.0 \pm 0.6$ & 4 & 6.6--6.9 \\
      \hline
      7j3b loose & $\geq$7 & $\geq$3 & $>$1000 & $>$400  & $5.1 \pm 1.5$ & 5 & 6.4--6.6 \\
      7j3b tight & $\geq$7 & $\geq$3 & $>$1500 & $>$400  & $0.9 \pm 0.5$ & 1 & 3.6--3.7 \\
      \hline
      \end{tabular}

\end{table*}

\subsection{Interpretation}
\label{sec:interpretation}

The results of the search can be interpreted by performing a maximum likelihood fit to the data in the signal regions.
The fit is carried out under either a background-only or a background+signal hypothesis.
The uncertainties in the modeling of the backgrounds, summarized in Section~\ref{sec:bkgds}, are inputs to the fitting procedure.
The likelihood is constructed as the product of Poisson probability density functions, one for each signal region, with
constraint terms that account for uncertainties in the background estimates and, if considered, the signal yields.
The result of the background-only fit, denoted as ``post-fit background,'' is given in Appendix~\ref{app:results}.
If the magnitude and correlation model of the uncertainties associated to the pre-fit estimates are properly assigned,
and the data are found to be in agreement with the estimates, then the fit has the effect of constraining the background and reducing the associated uncertainties.

\begin{figure*}[htbp]
  \centering
    \includegraphics[width=0.3\textwidth]{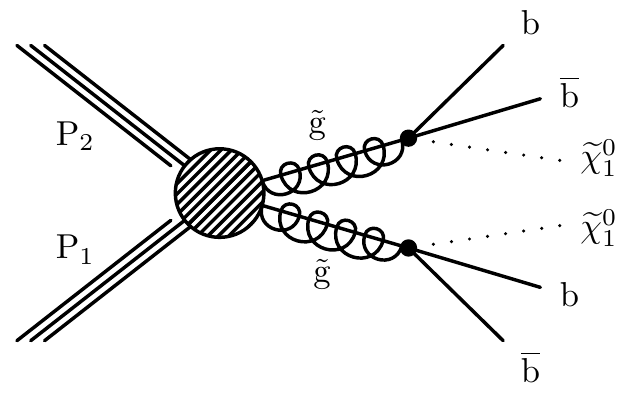}
    \includegraphics[width=0.3\textwidth]{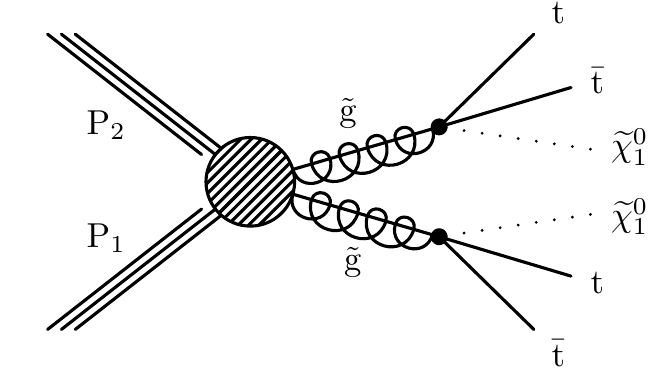}
    \includegraphics[width=0.3\textwidth]{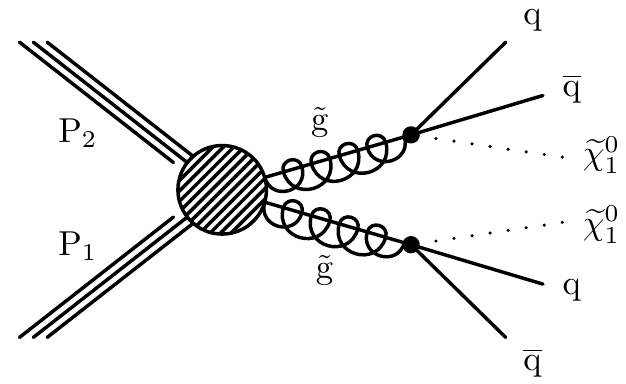} \\
    \includegraphics[width=0.3\textwidth]{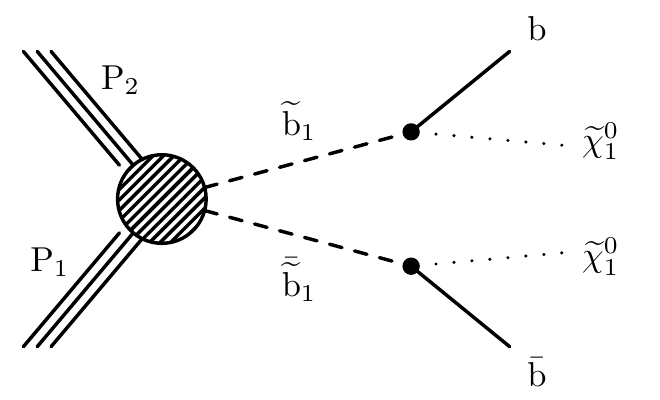}
    \includegraphics[width=0.3\textwidth]{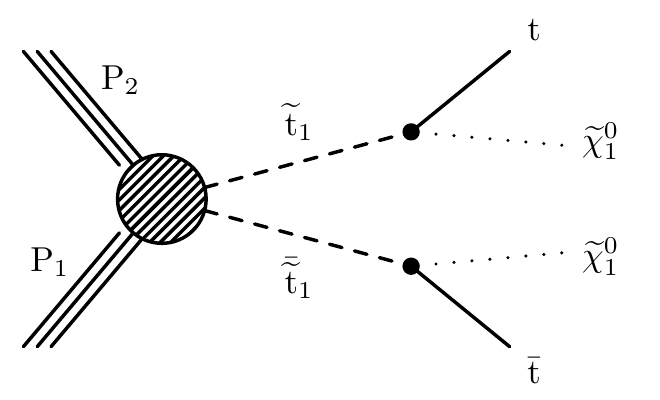}
    \includegraphics[width=0.3\textwidth]{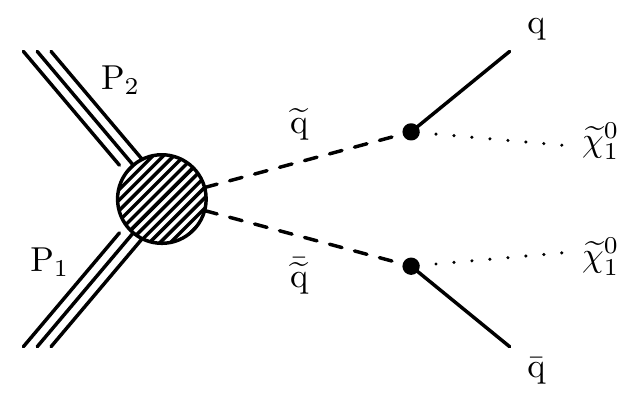} \\
    \includegraphics[width=0.3\textwidth]{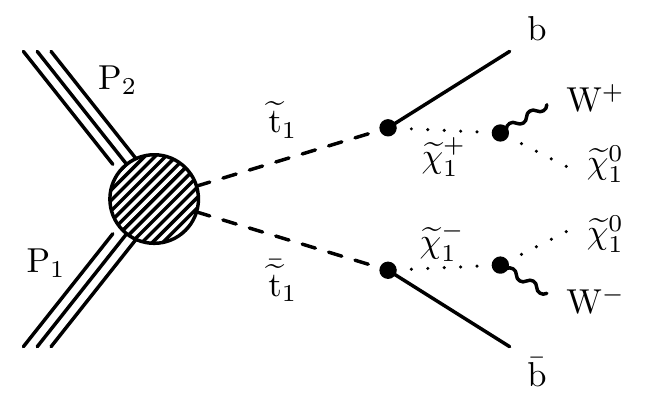}
    \includegraphics[width=0.3\textwidth]{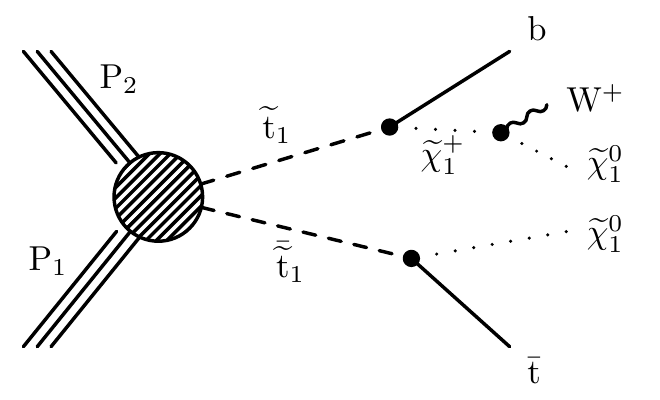}
    \includegraphics[width=0.3\textwidth]{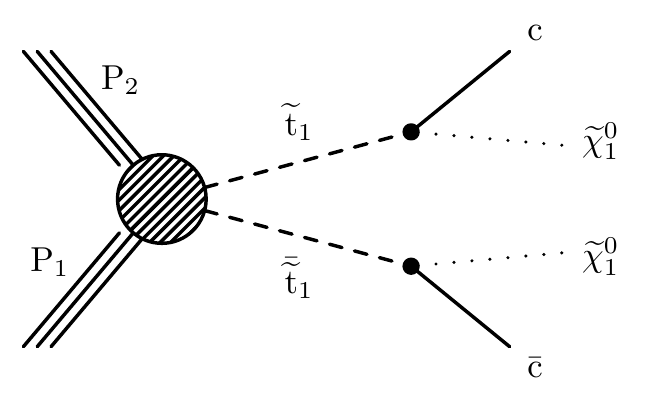} \\
    \caption{(Upper) Diagrams for the three scenarios of gluino-mediated bottom squark, top squark and light flavor squark production considered.
      (Middle) Diagrams for the direct production of bottom, top and light-flavor squark pairs.
      (Lower) Diagrams for three alternate scenarios of direct top squark production with different decay modes.
      For mixed decay scenarios, we assume a 50\% branching fraction
      for each decay
mode.}
    \label{fig:SMS_feynDiagrams}

\end{figure*}

The results of the search are used to constrain the simplified models of SUSY~\cite{SMSInterpretations} shown in Fig.~\ref{fig:SMS_feynDiagrams}.
For each scenario of gluino (squark) pair production, the simplified models assume that
all SUSY particles other than the gluino (squark)
and the lightest neutralino are too heavy to be produced directly, and that the gluino (squark) decays promptly.
The models assume that each gluino (squark) decays with a 100\% branching fraction into the decay products depicted in Fig.~\ref{fig:SMS_feynDiagrams}.
For models where the decays of the two squarks differ, we assume a 50\% branching fraction for each decay mode.
For the scenario of top squark pair production, the polarization of the top quark is model dependent
and is a function of the top-squark and neutralino mixing matrices.
To remain agnostic to a particular model realization, events are generated without polarization.
Signal cross sections are calculated at NLO+NLL order in $\alpha_{\mathrm{s}}$~\cite{bib-nlo-nll-01,bib-nlo-nll-02,bib-nlo-nll-03,bib-nlo-nll-04,bib-nlo-nll-05}.

Typical values of the uncertainties in the signal yield for the simplified models considered are listed in Table~\ref{tab:sig_systs}.
The sources of uncertainties and the methods used to evaluate their
effect on the interpretation are the same as those discussed in Ref.~\cite{MT2at13TeV}.
Uncertainties due to the luminosity~\cite{lumi2016}, ISR and pileup modeling, and \PQb tagging and lepton efficiencies are treated as correlated across search bins.
Remaining uncertainties are taken as uncorrelated.

\begin{figure*}[htbp]
  \centering
    \includegraphics[width=0.48\textwidth]{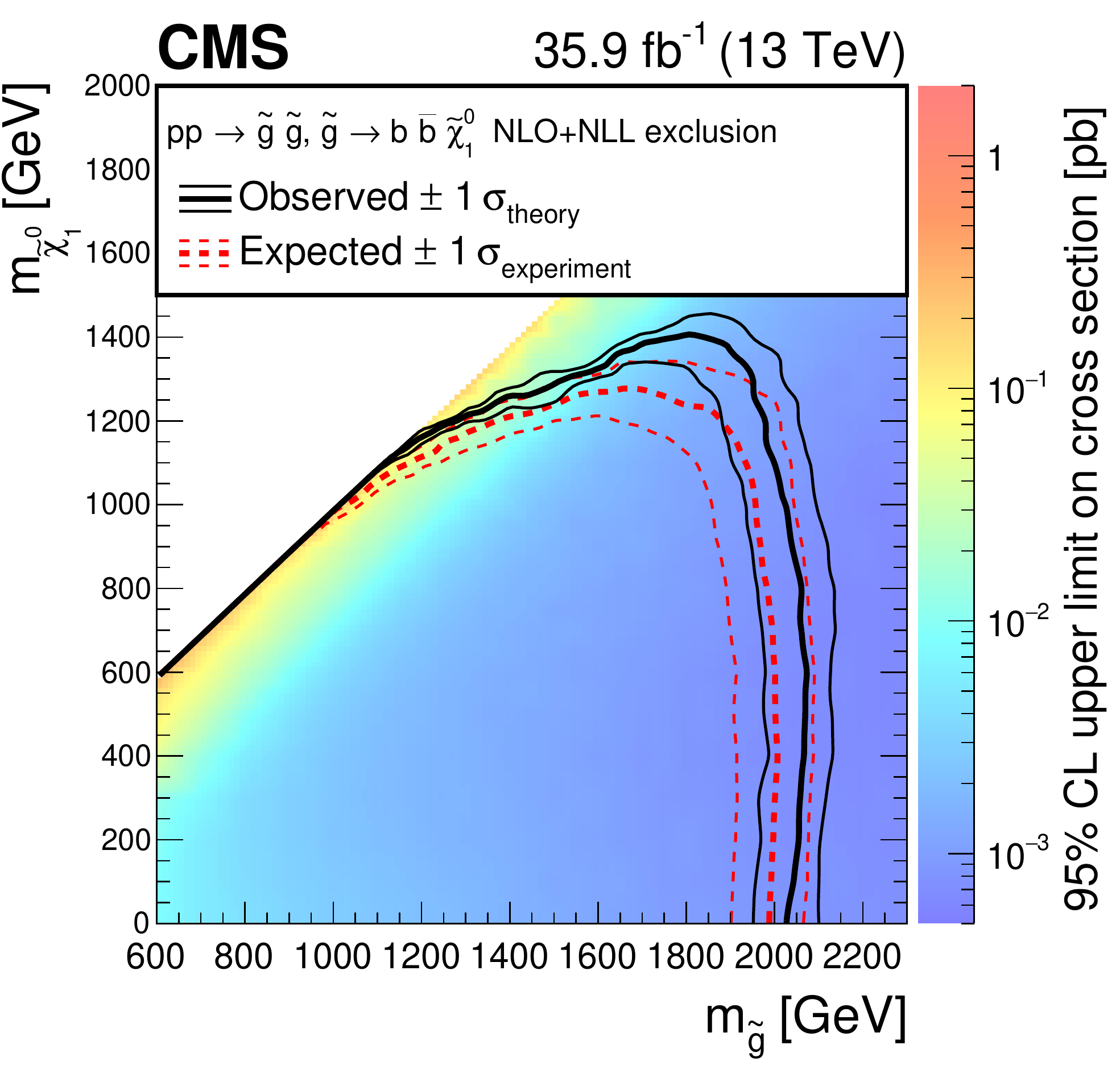}
    \includegraphics[width=0.48\textwidth]{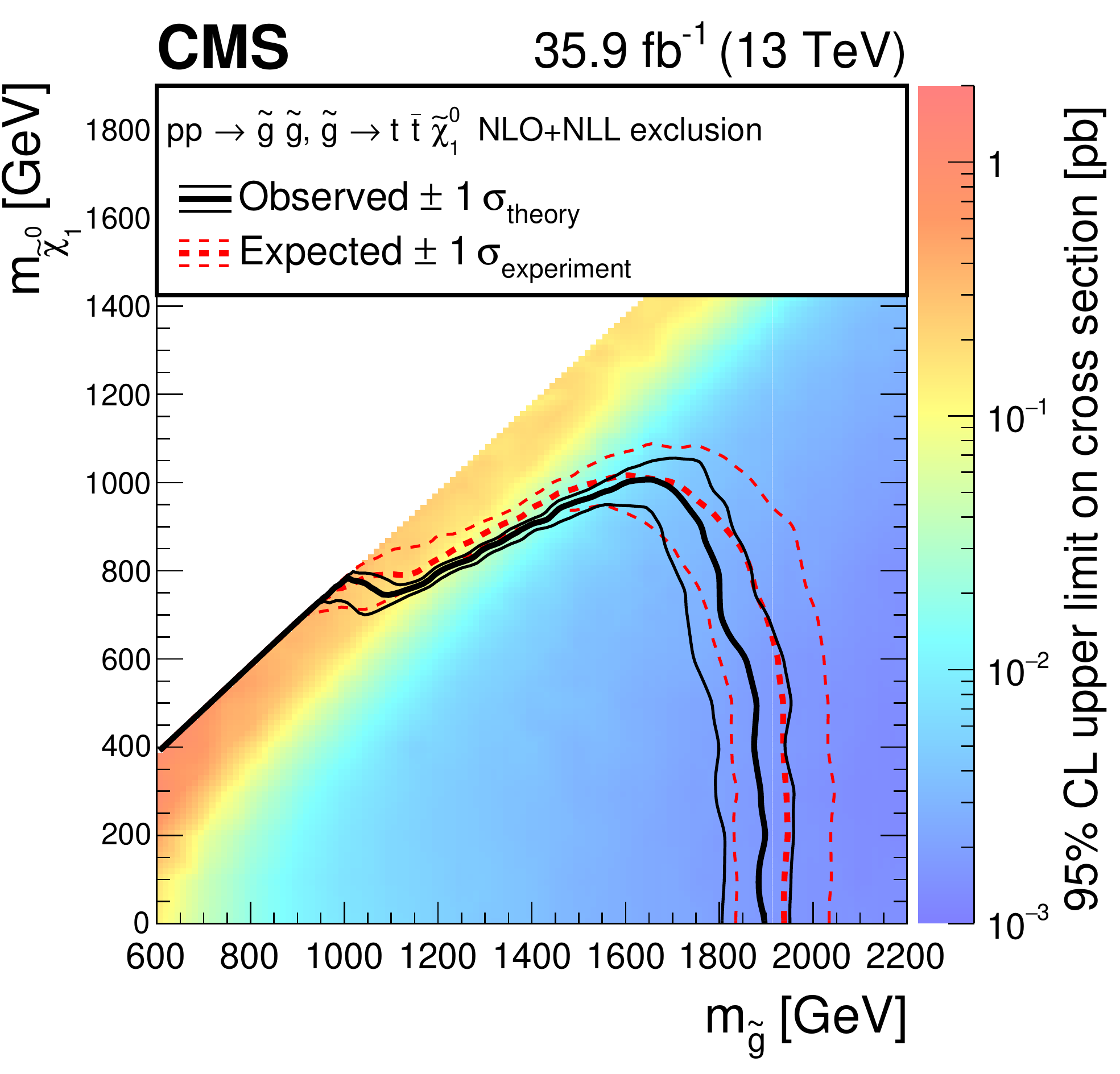}
    \includegraphics[width=0.48\textwidth]{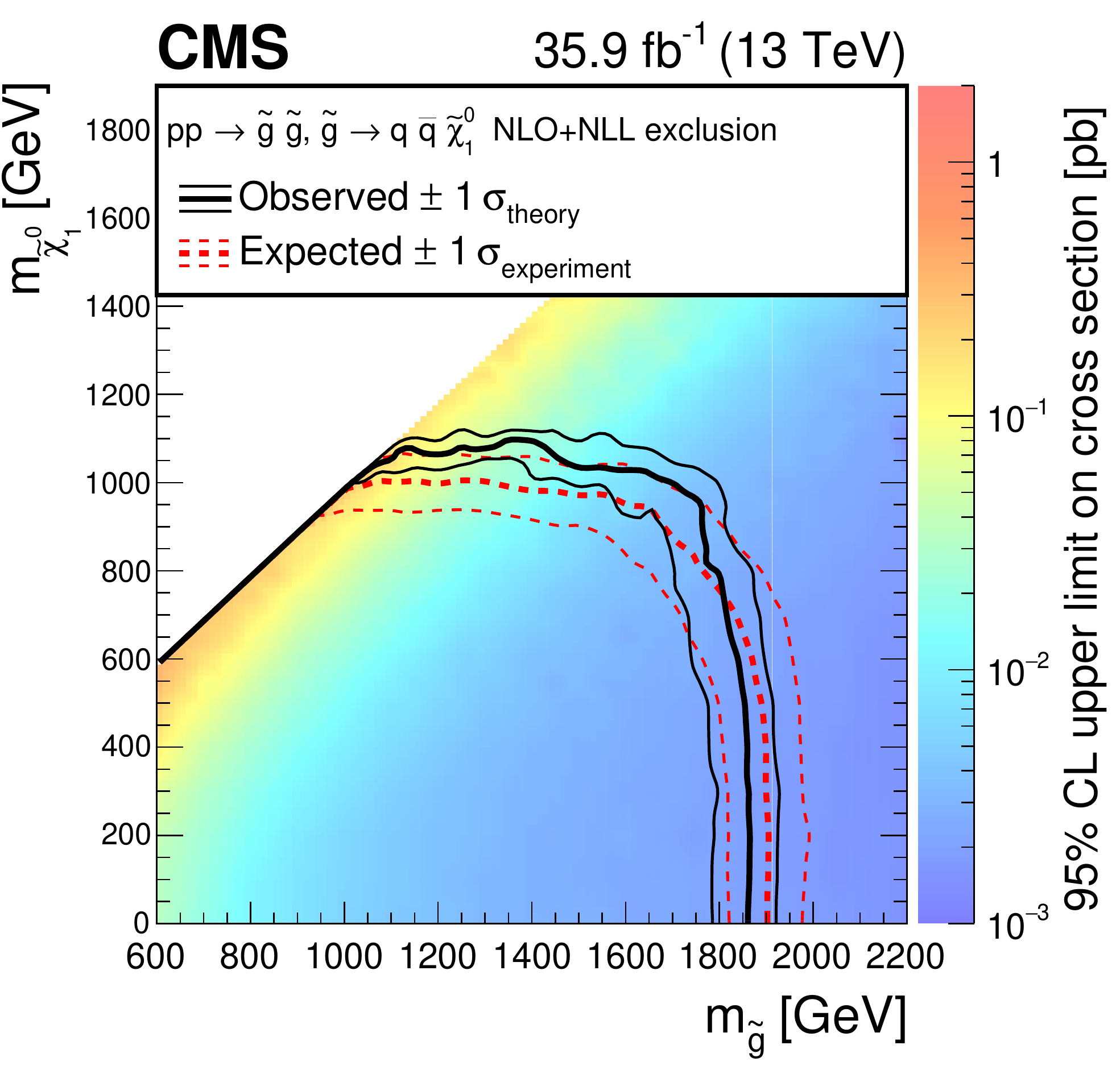}
    \caption{Exclusion limits at  95\% CL for gluino-mediated bottom squark production (above left),
    gluino-mediated top squark production (above right), and gluino-mediated light-flavor (\cPqu,\cPqd,\cPqs,\cPqc) squark production (below).
      The area enclosed by the thick black curve represents the observed exclusion region,
      while the dashed red lines indicate the expected limits and
      their $\pm$1 standard deviation ranges.
      The thin black lines show the effect of the theoretical
      uncertainties on the signal cross section.}
    \label{fig:t1x}
\end{figure*}

\begin{figure*}[htbp]
  \centering
    \includegraphics[width=0.48\textwidth]{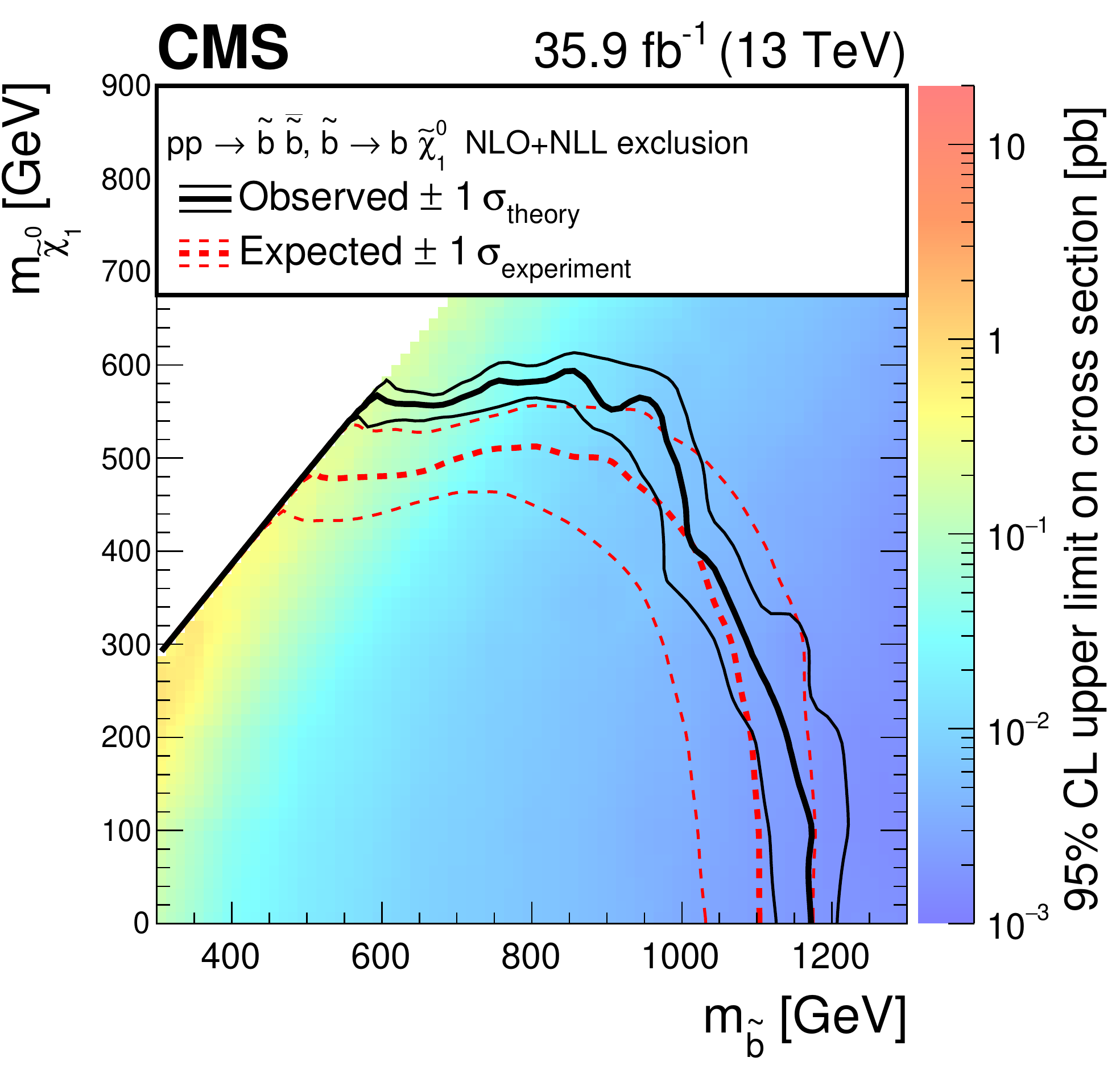}
    \includegraphics[width=0.48\textwidth]{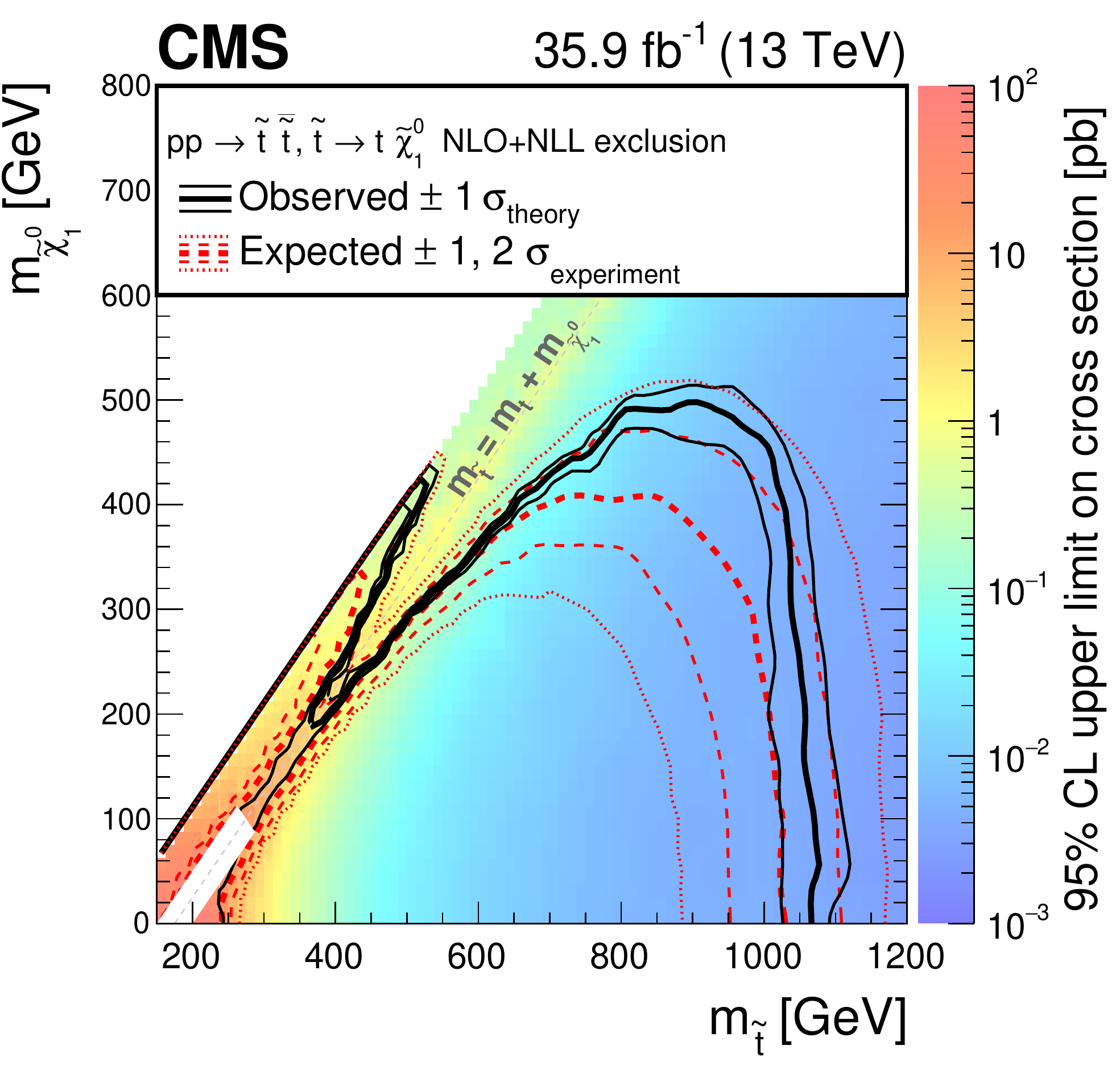}
    \includegraphics[width=0.48\textwidth]{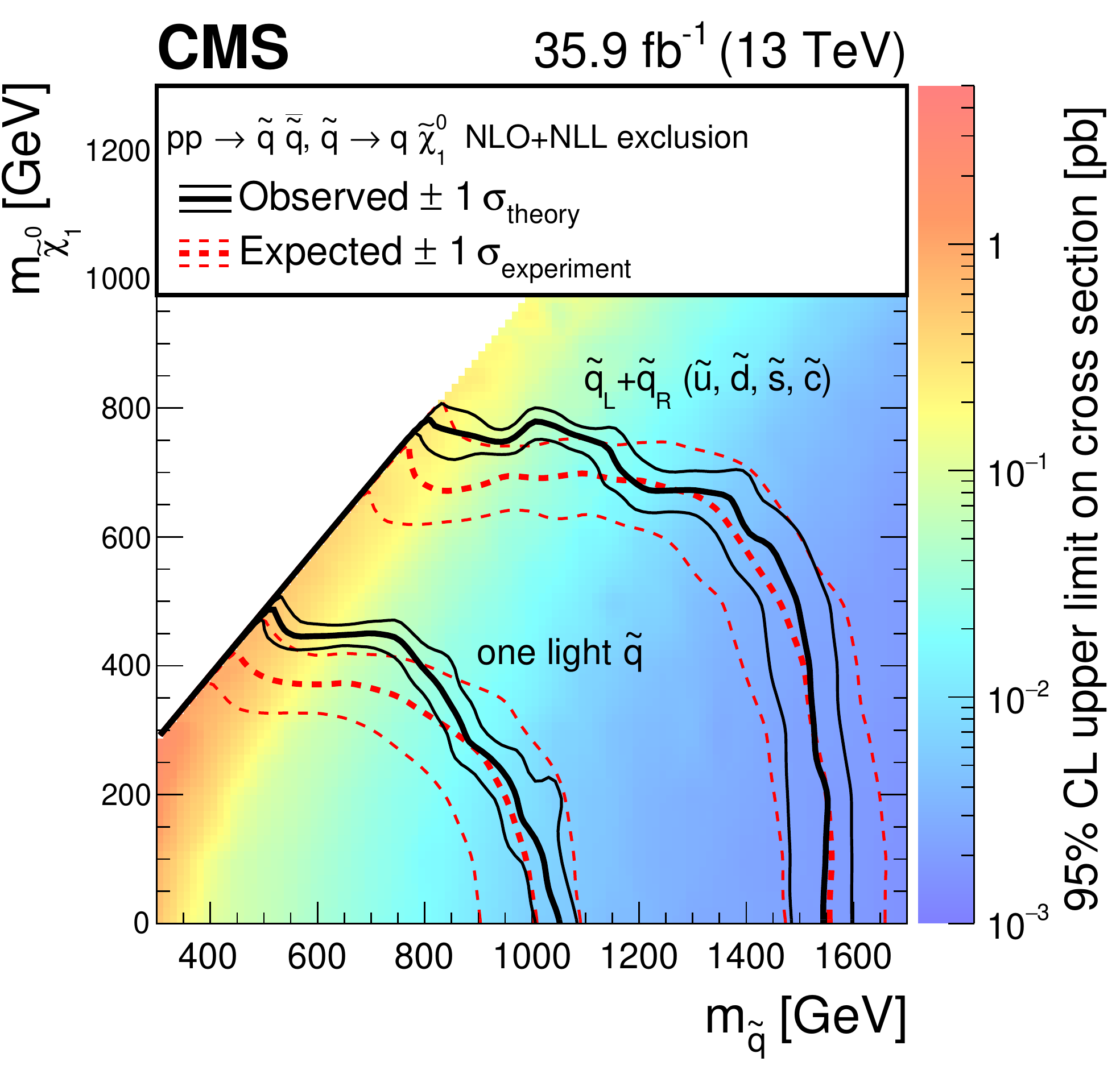}
    \caption{Exclusion limit at 95\% CL for bottom squark pair production (above left), top squark pair production (above right),
      and light-flavor squark pair production (below).
      The area enclosed by the thick black curve represents the observed exclusion region,
      while the dashed red lines indicate the expected limits and
      their $\pm$1 standard deviation ranges.
      For the top squark pair production plot, the $\pm$2 standard deviation
      ranges are also shown.
      The thin black lines show the effect of the theoretical
      uncertainties on the signal cross section.
      The white diagonal band in the upper right plot corresponds to the region
      $\abs{m_{\PSQt}-m_{\PQt}-m_{\lsp}}< 25\GeV$ and small $m_{\lsp}$. Here the efficiency of the selection
      is a strong function of $m_{\PSQt}-m_{\lsp}$, and as a result the precise
      determination of the cross section upper limit is uncertain
      because of the finite granularity of the available
      MC samples in this region of the ($m_{\PSQt}, m_{\lsp}$)  plane.}
    \label{fig:t2x}
\end{figure*}

\begin{figure*}[htbp]
  \centering
    \includegraphics[width=0.49\textwidth]{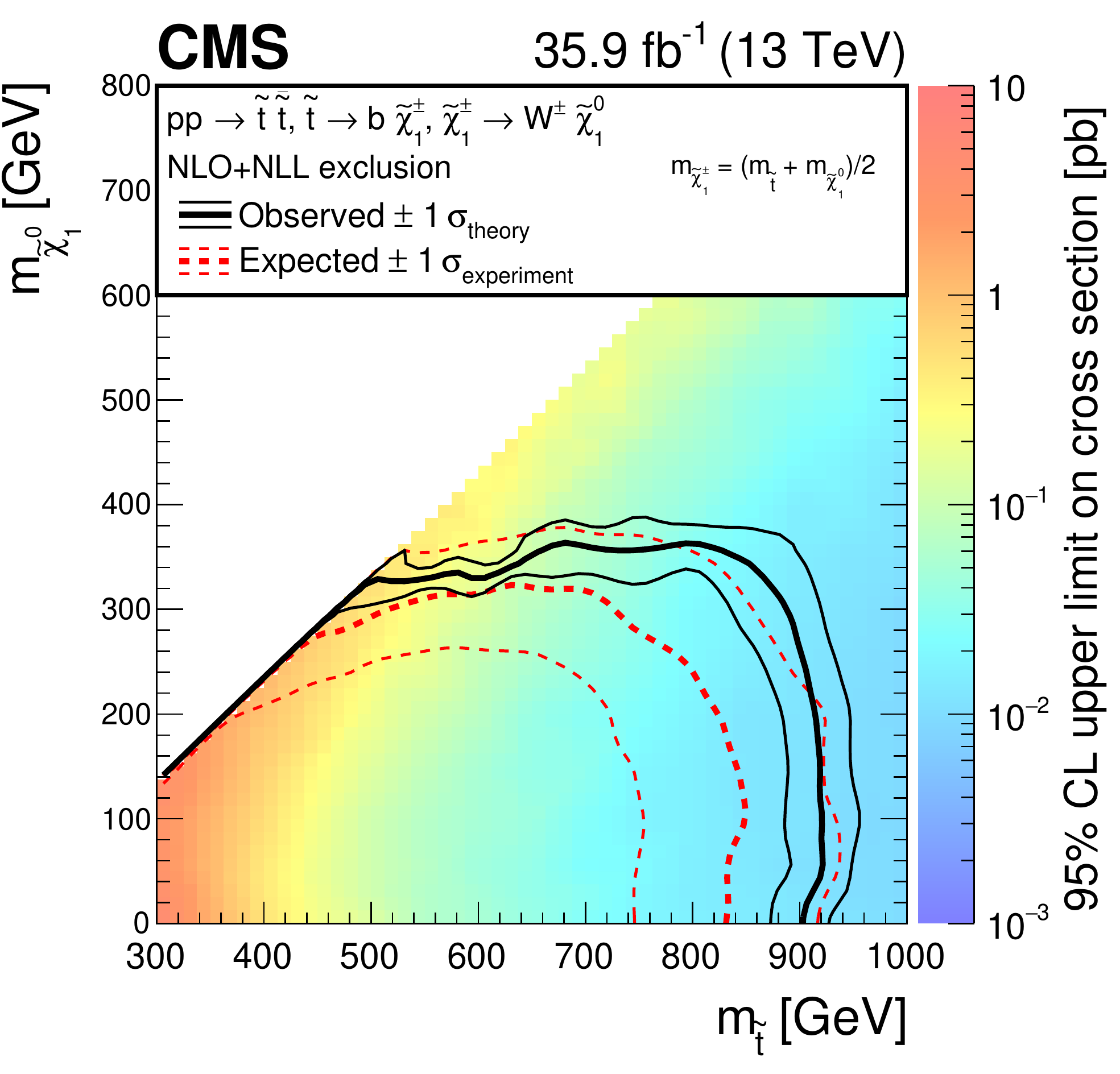}
    \includegraphics[width=0.49\textwidth]{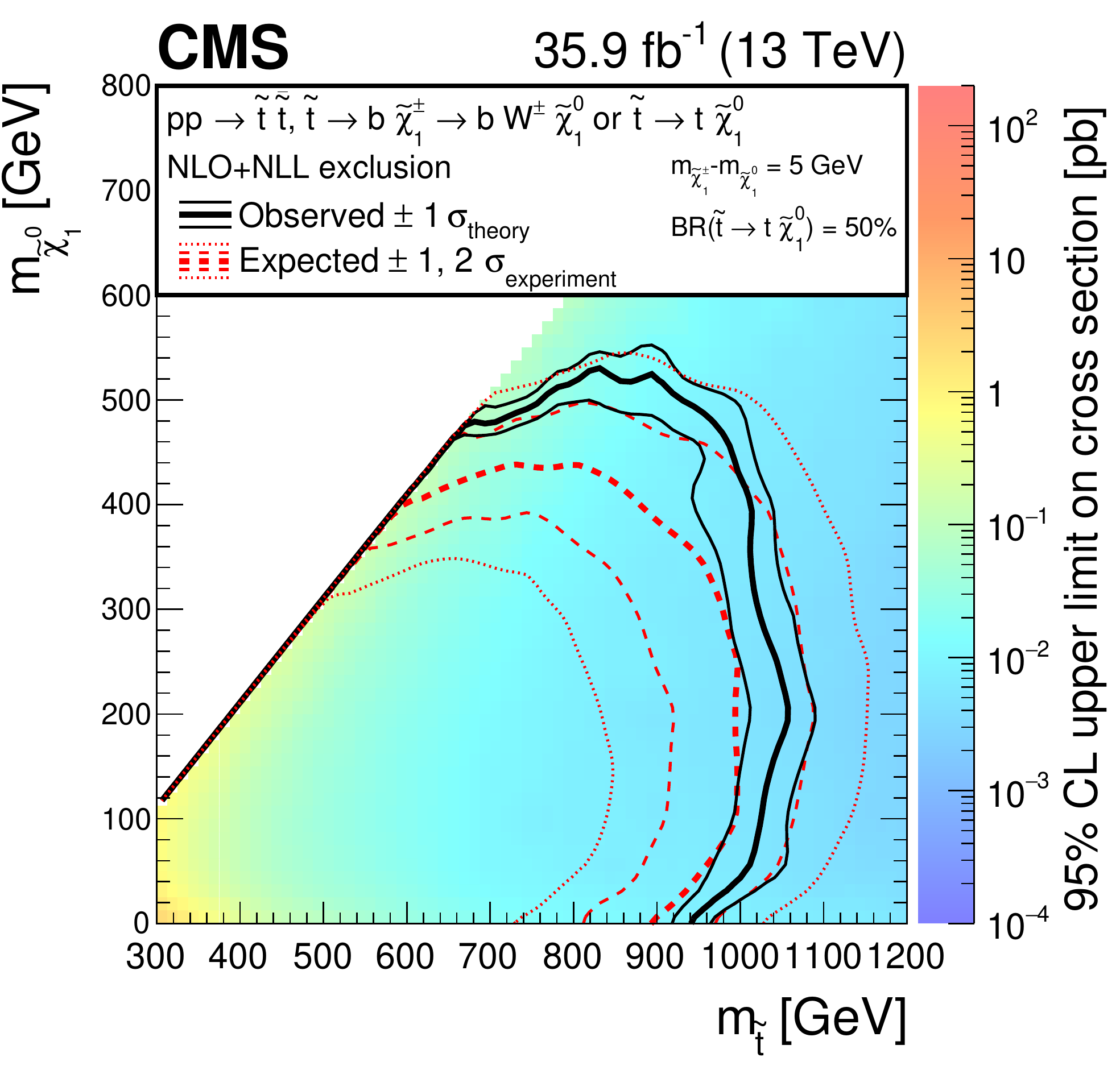}
    \includegraphics[width=0.49\textwidth]{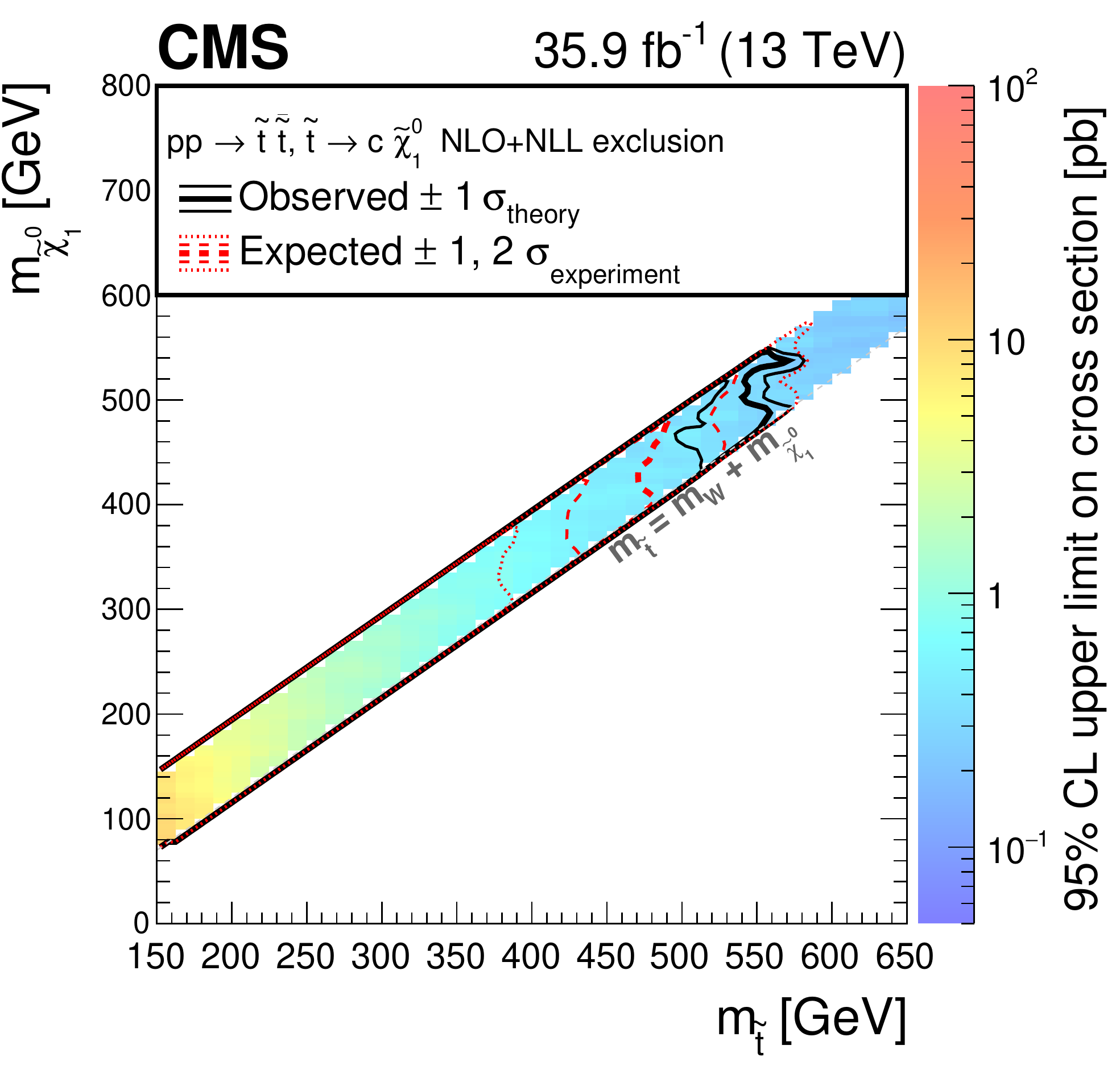}
    \caption{Exclusion limit at 95\% CL for top squark pair production for different decay modes of the top squark.
      For the scenario where $\Pp\Pp\to\PSQt\astop\to
      \PQb\PAQb\PSGcpmDo\PSGcmpDo$, $\PSGcpmDo\to\PW^{\pm} \PSGczDo$ (above left),
      the mass of the chargino is chosen to be half way in between the masses of the top squark and the neutralino.
      A mixed decay scenario (above right), $\Pp\Pp\to\PSQt\astop$ with equal branching fractions for the top squark decays $\PSQt\to\PQt\PSGczDo$
      and $\PSQt\to\PQb\PSGcpDo$, $\PSGcpDo\to\PW^{*+}\PSGczDo$,
      is also considered, with the chargino mass chosen such that
      $\Delta m\left(\PSGcpmDo,\PSGczDo\right) = 5\GeV$.  Finally, we
      also consider
      a compressed scenario (below) where
      $\Pp\Pp\to\PSQt\astop\to
      \PQc\PAQc\PSGczDo\PSGczDo$.
      The area enclosed by the thick black curve represents the observed exclusion region,
      while the dashed red lines indicate the expected limits and
      their $\pm$1 standard deviation ranges.
      The thin black lines show the effect of the theoretical
      uncertainties on the signal cross section.}
    \label{fig:stop_other}
\end{figure*}

Figure~\ref{fig:t1x} shows the exclusion limits at 95\% CL for
gluino-mediated bottom
squark,
top squark, and light-flavor squark production.
Exclusion limits at 95\% CL for the direct production of bottom,
top, and light-flavor squark pairs are shown in Fig.~\ref{fig:t2x}.
Direct production of top squarks for three alternate decay scenarios
are also
considered, and
exclusion limits at 95\% CL are shown in Fig.~\ref{fig:stop_other}.
Table~\ref{tab:lim} summarizes the limits on the masses of the SUSY particles excluded in the
simplified model scenarios considered.
These results extend the constraints on gluinos and squarks by about 300\GeV and on \lsp by 200\GeV with respect to those in Ref.~\cite{MT2at13TeV}.
The largest differences between the observed and expected limits are found for scenarios of top squark pair production
with moderate mass splittings and result from observed yields that are less than the expected background in topological regions with
\Ht between 575 and 1500~GeV, at least~7~jets, and either one or two b-tagged jets.

We note that
the 95\% CL upper limits on signal cross sections obtained using the
most sensitive super signal regions of Table~\ref{tab:ssr_def_yields}
are typically less stringent by a factor of $\sim$1.5--3 compared to
those obtained in the fully-binned analysis.
The full analysis performs better because of its larger signal acceptance and because it splits the events into bins with more favorable signal-to-background ratio.

\begin{table}[htb]
\topcaption{
Typical values of the systematic uncertainties as evaluated for the simplified models of SUSY used in the context of this search.
The high statistical uncertainty in the simulated signal sample corresponds to a small number of signal bins with low acceptance,
which are typically not among the most sensitive signal bins to that model point.
    \label{tab:sig_systs}}
\centering
\begin{tabular}{lc}
\hline
Source & Typical values [\%] \\
\hline
Integrated luminosity                & 2.5     \\
Limited size of MC samples                & 1--100  \\
Renormalization and factorization scales  & 5       \\
ISR modeling                            & 0--30   \\
b tagging efficiency, heavy flavors        & 0--40   \\
b tagging efficiency, light flavors        & 0--20   \\
Lepton efficiency                         & 0--20   \\
Jet energy scale                          & 5       \\
Fast simulation \Met\ modeling            & 0--5     \\
Fast simulation pileup modeling           & 4.6     \\
\hline
\end{tabular}
\end{table}

\begin{table*}[htb]
  \topcaption{Summary of 95\% CL observed exclusion limits on the masses of SUSY particles (sparticles) in different simplified model scenarios.
    The limit on the mass of the produced sparticle is quoted for a
    massless \lsp, while for the mass of the \lsp we quote the highest
    limit that is obtained.
    \label{tab:lim}}
\centering
\begin{tabular}{lrr}
\hline
Simplified & Limit on produced sparticle & Highest limit on the \\
model & mass [\GeVns{}] for $m_{\lsp}=0\GeV$ & \lsp mass [\GeVns{}] \\
\hline\hline
Direct squark production: & & \\
Bottom squark & 1175 & 590 \\
Top squark & 1070 & 550 \\
Single light squark & 1050 & 475 \\
Eight degenerate light squarks & 1550 & 775 \\
\hline
Gluino-mediated production: & & \\
$\gluino\to \bbbar\lsp$ & 2025 & 1400 \\
$\gluino\to \ttbar\lsp$ & 1900 & 1010 \\
$\gluino\to \qqbar\lsp$ & 1860 & 1100 \\
\hline
\end{tabular}
\end{table*}

\section{Summary}
\label{sec:conclusions}

This paper presents the results of a search for new phenomena using events with jets and large \mttwo.
Results are based on a \Lint data sample of proton-proton collisions at $\sqrt{s} =13\TeV$ collected in 2016 with the CMS detector.
No significant deviations from the standard model expectations are observed.
The results are interpreted as limits on the production of new, massive colored particles in simplified models of supersymmetry.
This search probes gluino masses up to 2025\GeV and \lsp masses up to 1400\GeV.
Constraints are also obtained on the pair production of light-flavor, bottom, and top squarks, probing masses up
to 1550, 1175, and 1070\GeV, respectively, and \lsp masses up to 775, 590, and 550\GeV in each scenario.

\begin{acknowledgments}
We congratulate our colleagues in the CERN accelerator departments for the excellent performance of the LHC and thank the technical and administrative staffs at CERN and at other CMS institutes for their contributions to the success of the CMS effort. In addition, we gratefully acknowledge the computing centers and personnel of the Worldwide LHC Computing Grid for delivering so effectively the computing infrastructure essential to our analyses. Finally, we acknowledge the enduring support for the construction and operation of the LHC and the CMS detector provided by the following funding agencies: BMWFW and FWF (Austria); FNRS and FWO (Belgium); CNPq, CAPES, FAPERJ, and FAPESP (Brazil); MES (Bulgaria); CERN; CAS, MoST, and NSFC (China); COLCIENCIAS (Colombia); MSES and CSF (Croatia); RPF (Cyprus); SENESCYT (Ecuador); MoER, ERC IUT, and ERDF (Estonia); Academy of Finland, MEC, and HIP (Finland); CEA and CNRS/IN2P3 (France); BMBF, DFG, and HGF (Germany); GSRT (Greece); OTKA and NIH (Hungary); DAE and DST (India); IPM (Iran); SFI (Ireland); INFN (Italy); MSIP and NRF (Republic of Korea); LAS (Lithuania); MOE and UM (Malaysia); BUAP, CINVESTAV, CONACYT, LNS, SEP, and UASLP-FAI (Mexico); MBIE (New Zealand); PAEC (Pakistan); MSHE and NSC (Poland); FCT (Portugal); JINR (Dubna); MON, RosAtom, RAS, RFBR and RAEP (Russia); MESTD (Serbia); SEIDI, CPAN, PCTI and FEDER (Spain); Swiss Funding Agencies (Switzerland); MST (Taipei); ThEPCenter, IPST, STAR, and NSTDA (Thailand); TUBITAK and TAEK (Turkey); NASU and SFFR (Ukraine); STFC (United Kingdom); DOE and NSF (USA).

\hyphenation{Rachada-pisek} Individuals have received support from the Marie-Curie program and the European Research Council and EPLANET (European Union); the Leventis Foundation; the A. P. Sloan Foundation; the Alexander von Humboldt Foundation; the Belgian Federal Science Policy Office; the Fonds pour la Formation \`a la Recherche dans l'Industrie et dans l'Agriculture (FRIA-Belgium); the Agentschap voor Innovatie door Wetenschap en Technologie (IWT-Belgium); the Ministry of Education, Youth and Sports (MEYS) of the Czech Republic; the Council of Science and Industrial Research, India; the HOMING PLUS program of the Foundation for Polish Science, cofinanced from European Union, Regional Development Fund, the Mobility Plus program of the Ministry of Science and Higher Education, the National Science Center (Poland), contracts Harmonia 2014/14/M/ST2/00428, Opus 2014/13/B/ST2/02543, 2014/15/B/ST2/03998, and 2015/19/B/ST2/02861, Sonata-bis 2012/07/E/ST2/01406; the National Priorities Research Program by Qatar National Research Fund; the Programa Clar\'in-COFUND del Principado de Asturias; the Thalis and Aristeia programs cofinanced by EU-ESF and the Greek NSRF; the Rachadapisek Sompot Fund for Postdoctoral Fellowship, Chulalongkorn University and the Chulalongkorn Academic into Its 2nd Century Project Advancement Project (Thailand); and the Welch Foundation, contract C-1845.
\end{acknowledgments}
\clearpage
\bibliography{auto_generated}

\clearpage
\appendix

\section{Definition of search regions}
\label{app:srs}

The 213 exclusive search regions are defined in Tables~\ref{tab:sr1}--\ref{tab:sr3}.

\begin{table}[htb]
  \centering
    \topcaption{\label{tab:sr1} Summary of signal regions for the monojet selection.}
    \begin{tabular}{cl}
      \hline
      \nbtags & Jet \pt binning [\GeVns{}] \\
      \hline
      0 & [250, 350, 450, 575, 700, 1000, 1200, $\infty$) \\
      $\geq1$ & [250, 350, 450, 575, 700, $\infty$) \\
      \hline
    \end{tabular}

\end{table}

\begin{table*}[htb]
  \centering
    \topcaption{\label{tab:sr2} The \mttwo binning in each topological region of the multi-jet search regions,
      for the very low, low and medium \Ht regions.}
    \begin{tabular}{ccl}
      \hline
      \Ht range [\GeVns{}] & Jet multiplicities & \mttwo binning [\GeVns{}] \\
      \hline
          [ 250, 450 ] & $2-3$j, $  0$b  &  [ 200, 300, 400,  $\infty$  ) \\
          & $2-3$j, $  1$b  &  [ 200, 300, 400,  $\infty$  ) \\
          & $2-3$j, $  2$b  &  [ 200, 300, 400,  $\infty$  ) \\
          & $\geq4$j, $  0$b  &  [ 200, 300, 400,  $\infty$  ) \\
          & $\geq4$j, $  1$b  &  [ 200, 300, 400,  $\infty$  ) \\
          & $\geq4$j, $  2$b  &  [ 200, 300, 400,  $\infty$  ) \\
          & $\geq2$j, $  \geq3$b  &  [ 200, 300, 400,  $\infty$  ) \\
          \hline
          [ 450, 575 ] & $2-3$j, $  0$b  &  [ 200, 300, 400, 500,  $\infty$  ) \\
          & $2-3$j, $  1$b  &  [ 200, 300, 400, 500,  $\infty$  ) \\
          & $2-3$j, $  2$b  &  [ 200, 300, 400, 500,  $\infty$  ) \\
          & $4-6$j, $  0$b  &  [ 200, 300, 400, 500,  $\infty$  ) \\
          & $4-6$j, $  1$b  &  [ 200, 300, 400, 500,  $\infty$  ) \\
          & $4-6$j, $  2$b  &  [ 200, 300, 400, 500,  $\infty$  ) \\
          & $\geq7$j, $  0$b  &  [ 200, 300, 400,  $\infty$  ) \\
          & $\geq7$j, $  1$b  &  [ 200, 300, 400,  $\infty$  ) \\
          & $\geq7$j, $  2$b  &  [ 200, 300, 400,  $\infty$  ) \\
          & $2-6$j, $  \geq3$b  &  [ 200, 300, 400, 500,  $\infty$  ) \\
          & $\geq7$j, $  \geq3$b  &  [ 200, 300, 400,  $\infty$  ) \\
          \hline
          [ 575, 1000 ] & $2-3$j, $  0$b  &  [ 200, 300, 400, 600, 800,  $\infty$  ) \\
          & $2-3$j, $  1$b  &  [ 200, 300, 400, 600, 800,  $\infty$  ) \\
          & $2-3$j, $  2$b  &  [ 200, 300, 400, 600, 800,  $\infty$  ) \\
          & $4-6$j, $  0$b  &  [ 200, 300, 400, 600, 800,  $\infty$  ) \\
          & $4-6$j, $  1$b  &  [ 200, 300, 400, 600, 800,  $\infty$  ) \\
          & $4-6$j, $  2$b  &  [ 200, 300, 400, 600, 800,  $\infty$  ) \\
          & $\geq7$j, $  0$b  &  [ 200, 300, 400, 600, 800,  $\infty$  ) \\
          & $\geq7$j, $  1$b  &  [ 200, 300, 400, 600,  $\infty$  ) \\
          & $\geq7$j, $  2$b  &  [ 200, 300, 400, 600,  $\infty$  ) \\
          & $2-6$j, $  \geq3$b  &  [ 200, 300, 400, 600,  $\infty$  ) \\
          & $\geq7$j, $  \geq3$b  &  [ 200, 300, 400, 600,  $\infty$  ) \\
          \hline
    \end{tabular}

\end{table*}

\begin{table*}[htb]
  \centering
    \topcaption{\label{tab:sr3} The \mttwo binning in each topological region of the multijet search regions,
      for the high- and extreme-\Ht regions.}
    \begin{tabular}{ccl}
      \hline
      \Ht range [\GeVns{}] & Jet multiplicities & \mttwo binning [\GeVns{}] \\
      \hline
          [ 1000, 1500 ] & $2-3$j, $  0$b  &  [ 200, 400, 600, 800, 1000, 1200,  $\infty$  ) \\
          & $2-3$j, $  1$b  &  [ 200, 400, 600, 800, 1000, 1200,  $\infty$  ) \\
          & $2-3$j, $  2$b  &  [ 200, 400, 600, 800, 1000,  $\infty$  ) \\
          & $4-6$j, $  0$b  &  [ 200, 400, 600, 800, 1000, 1200,  $\infty$  ) \\
          & $4-6$j, $  1$b  &  [ 200, 400, 600, 800, 1000, 1200,  $\infty$  ) \\
          & $4-6$j, $  2$b  &  [ 200, 400, 600, 800, 1000,  $\infty$  ) \\
          & $\geq7$j, $  0$b  &  [ 200, 400, 600, 800, 1000,  $\infty$  ) \\
          & $\geq7$j, $  1$b  &  [ 200, 400, 600, 800,  $\infty$  ) \\
          & $\geq7$j, $  2$b  &  [ 200, 400, 600, 800,  $\infty$  ) \\
          & $2-6$j, $  \geq3$b  &  [ 200, 400, 600,  $\infty$  ) \\
          & $\geq7$j, $  \geq3$b  &  [ 200, 400, 600,  $\infty$  ) \\
          \hline
          [ 1500, $\infty$  ) & $2-3$j, $  0$b  &  [ 400, 600, 800, 1000, 1400,  $\infty$  ) \\
          & $2-3$j, $  1$b  &  [ 400, 600, 800, 1000,  $\infty$  ) \\
          & $2-3$j, $  2$b  &  [ 400,  $\infty$  ) \\
          & $4-6$j, $  0$b  &  [ 400, 600, 800, 1000, 1400,  $\infty$  ) \\
          & $4-6$j, $  1$b  &  [ 400, 600, 800, 1000, 1400,  $\infty$  ) \\
          & $4-6$j, $  2$b  &  [ 400, 600, 800,  $\infty$  ) \\
          & $\geq7$j, $  0$b  &  [ 400, 600, 800, 1000,  $\infty$  ) \\
          & $\geq7$j, $  1$b  &  [ 400, 600, 800,  $\infty$  ) \\
          & $\geq7$j, $  2$b  &  [ 400, 600, 800,  $\infty$  ) \\
          & $2-6$j, $  \geq3$b  &  [ 400, 600, $\infty$  ) \\
          & $\geq7$j, $  \geq3$b  &  [ 400, $\infty$  ) \\
      \hline
    \end{tabular}
\end{table*}
\clearpage
\section{Detailed results}
\label{app:results}

\begin{figure*}[htb]
  \centering
    \includegraphics[width=0.79\textwidth]{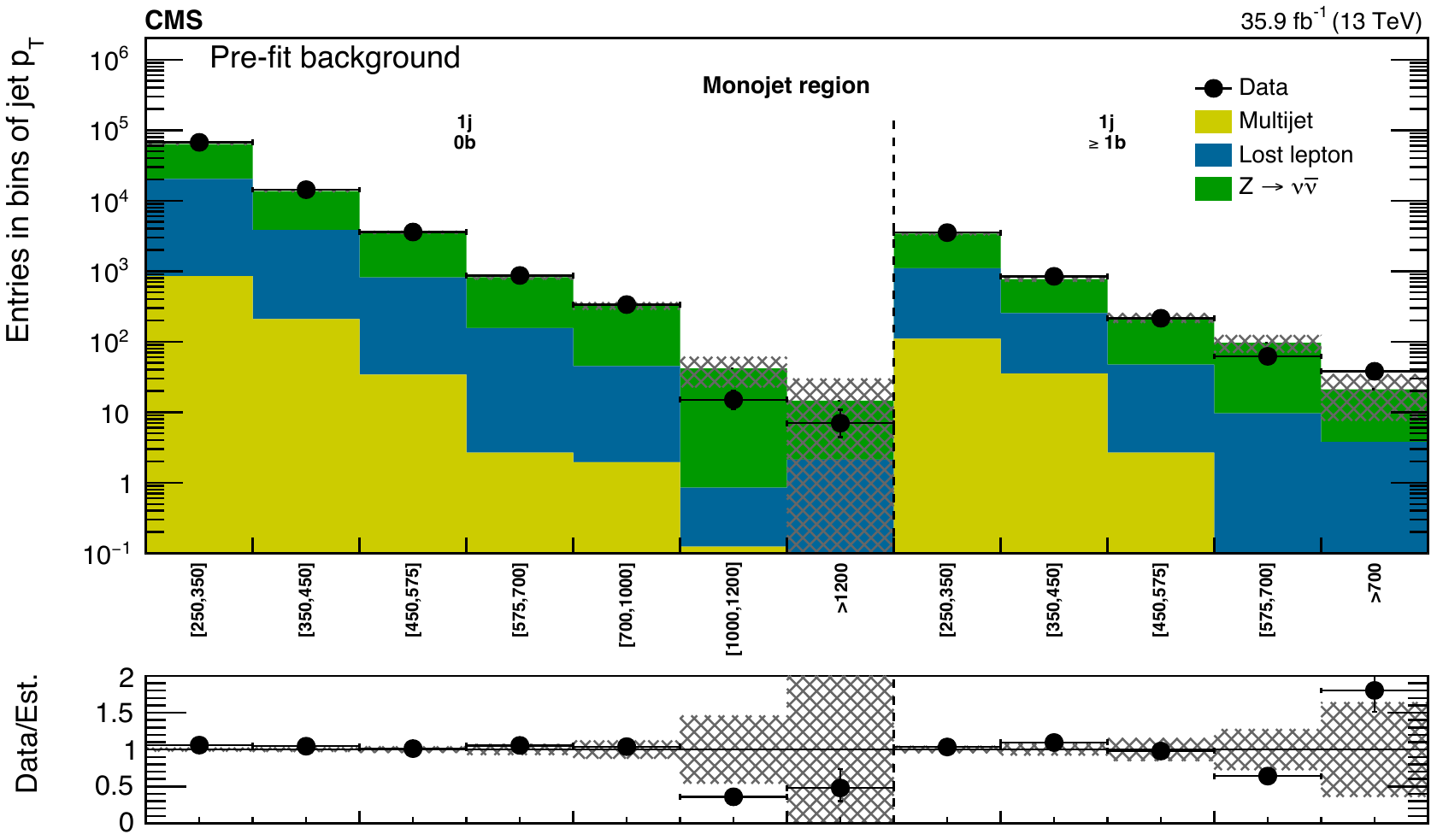}\\
    \includegraphics[width=0.79\textwidth]{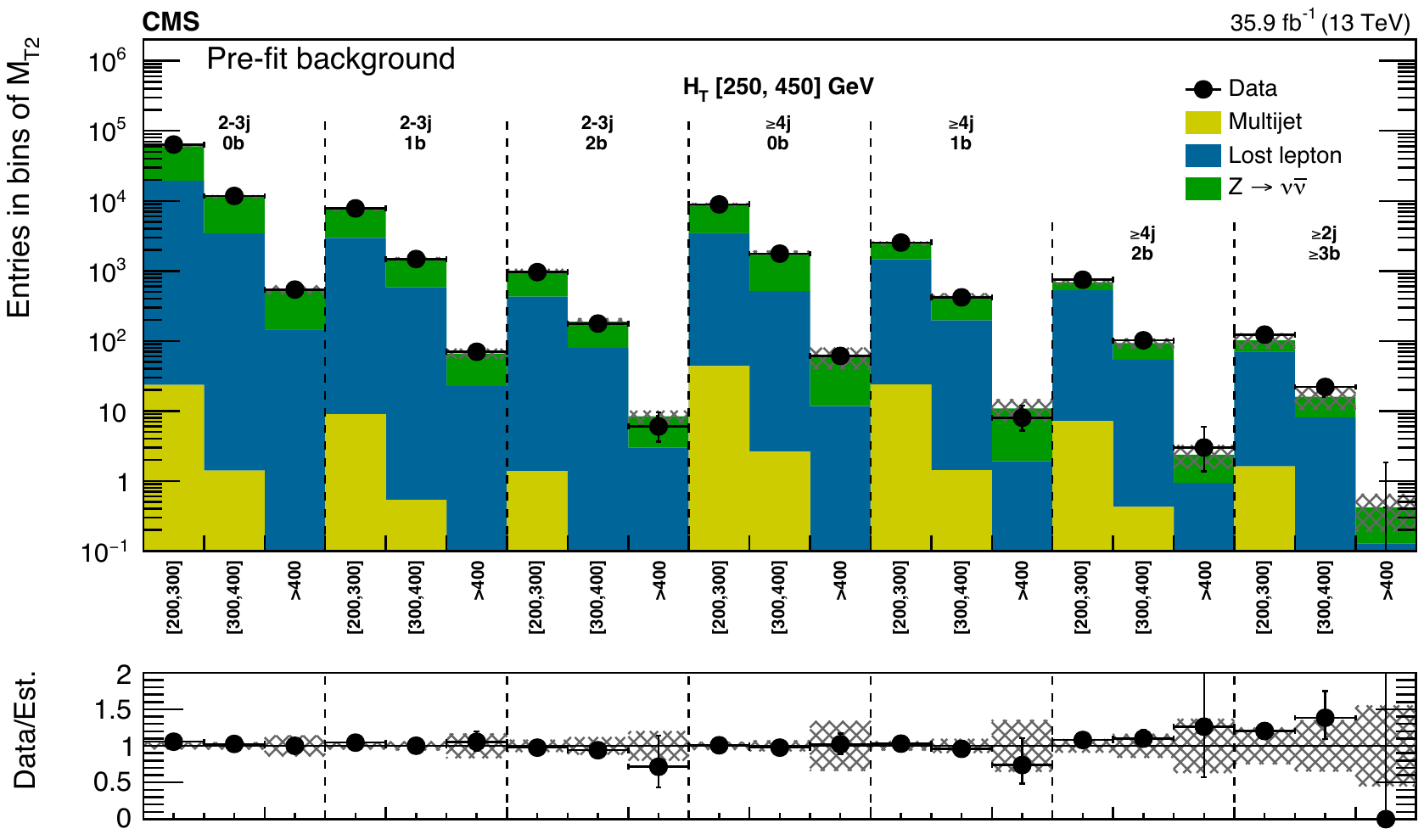}\\
    \caption{(Upper) Comparison of the estimated background and observed data events in each signal bin in the monojet region. On the $x$-axis, the \ptj binning is shown in units of \GeV. Hatched bands represent the full uncertainty in the background estimate.
    (Lower) Same for the very low \Ht region.  On the $x$-axis, the \mttwo binning is shown in units of \GeV.}
    \label{fig:otherResults1}
\end{figure*}

\begin{figure*}[htb]
  \centering
    \includegraphics[width=0.72\textwidth]{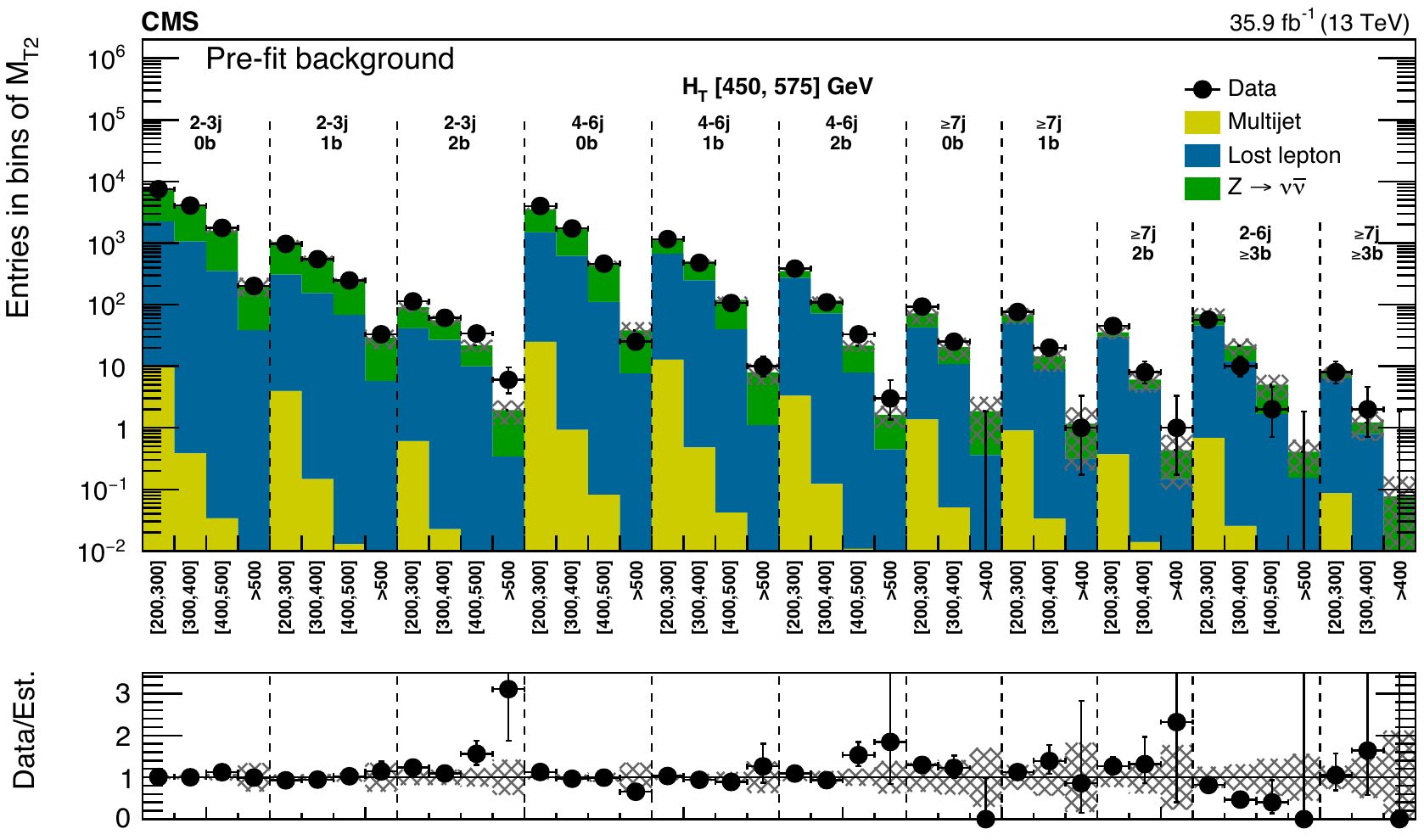}\\
    \includegraphics[width=0.72\textwidth]{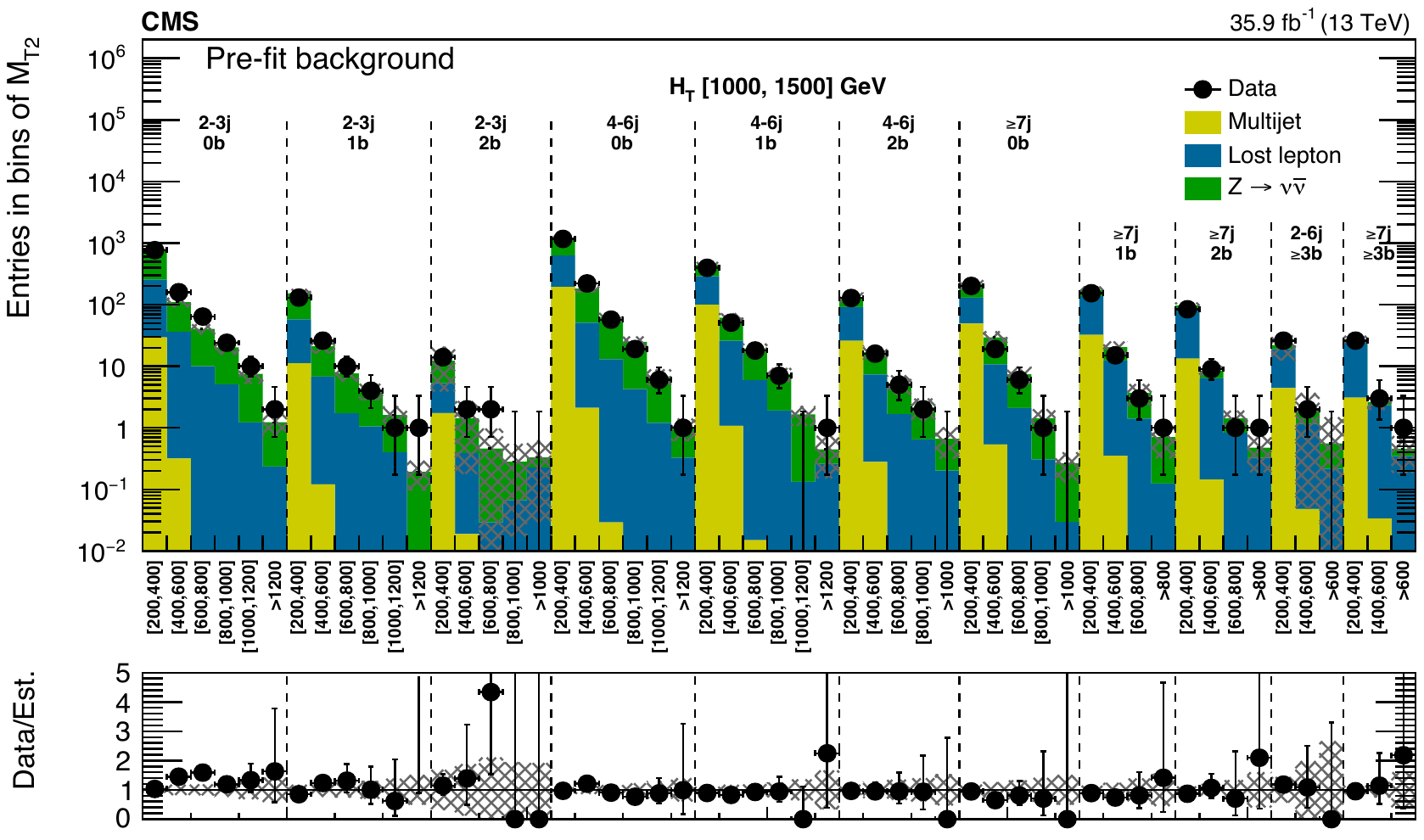}\\
    \includegraphics[width=0.72\textwidth]{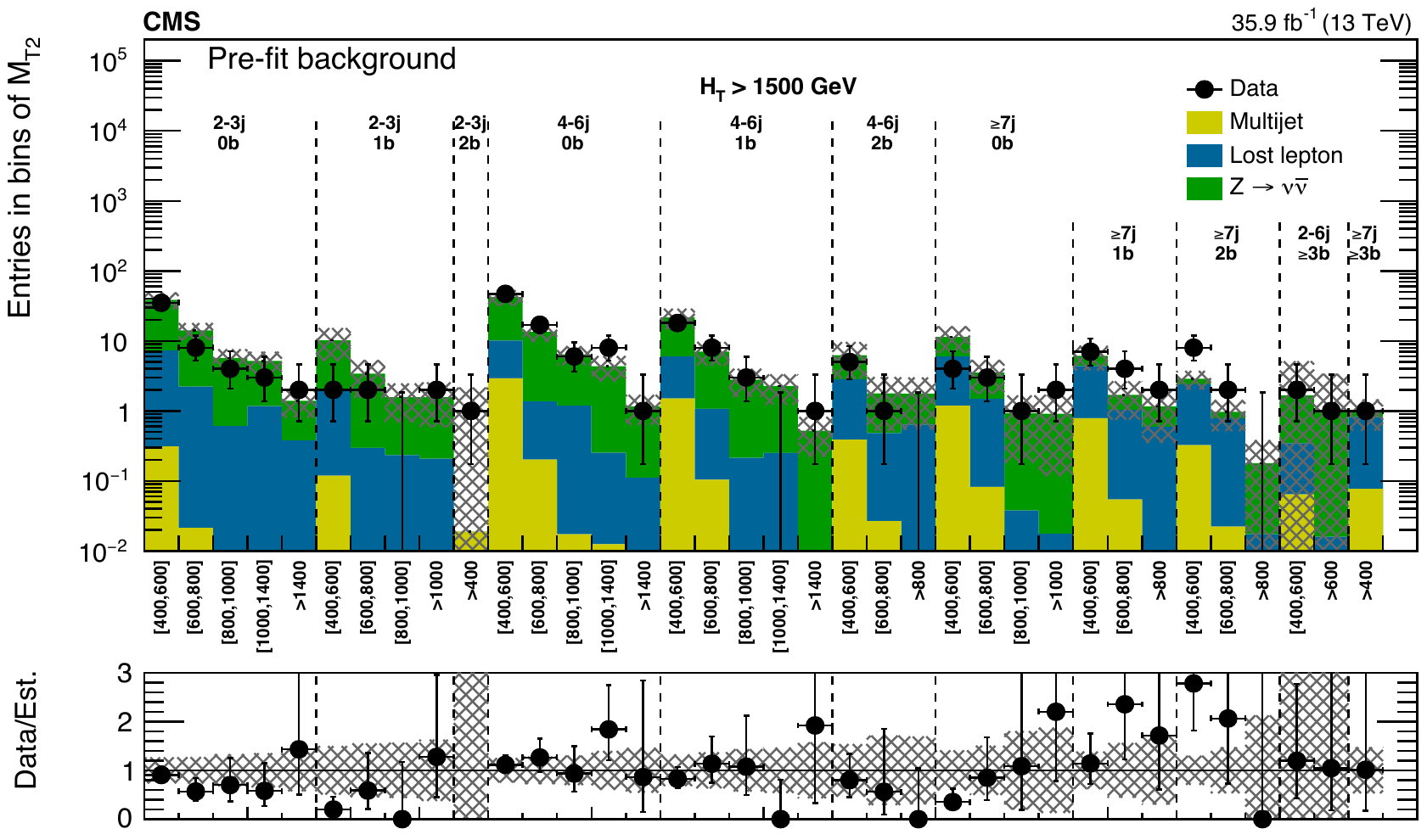}\\
    \caption{(Upper) Comparison of the estimated background and observed data events in each signal bin in the low-\Ht region.  Hatched bands represent the full uncertainty in the background estimate.
    Same for the high- (middle) and extreme- (lower) \Ht regions. On the $x$-axis, the \mttwo binning is shown in units of \GeV.
    For the extreme-\Ht region, the last bin is left empty for visualization purposes.}
    \label{fig:otherResults2}
\end{figure*}

\begin{figure*}[htb]
  \centering
    \includegraphics[width=\textwidth]{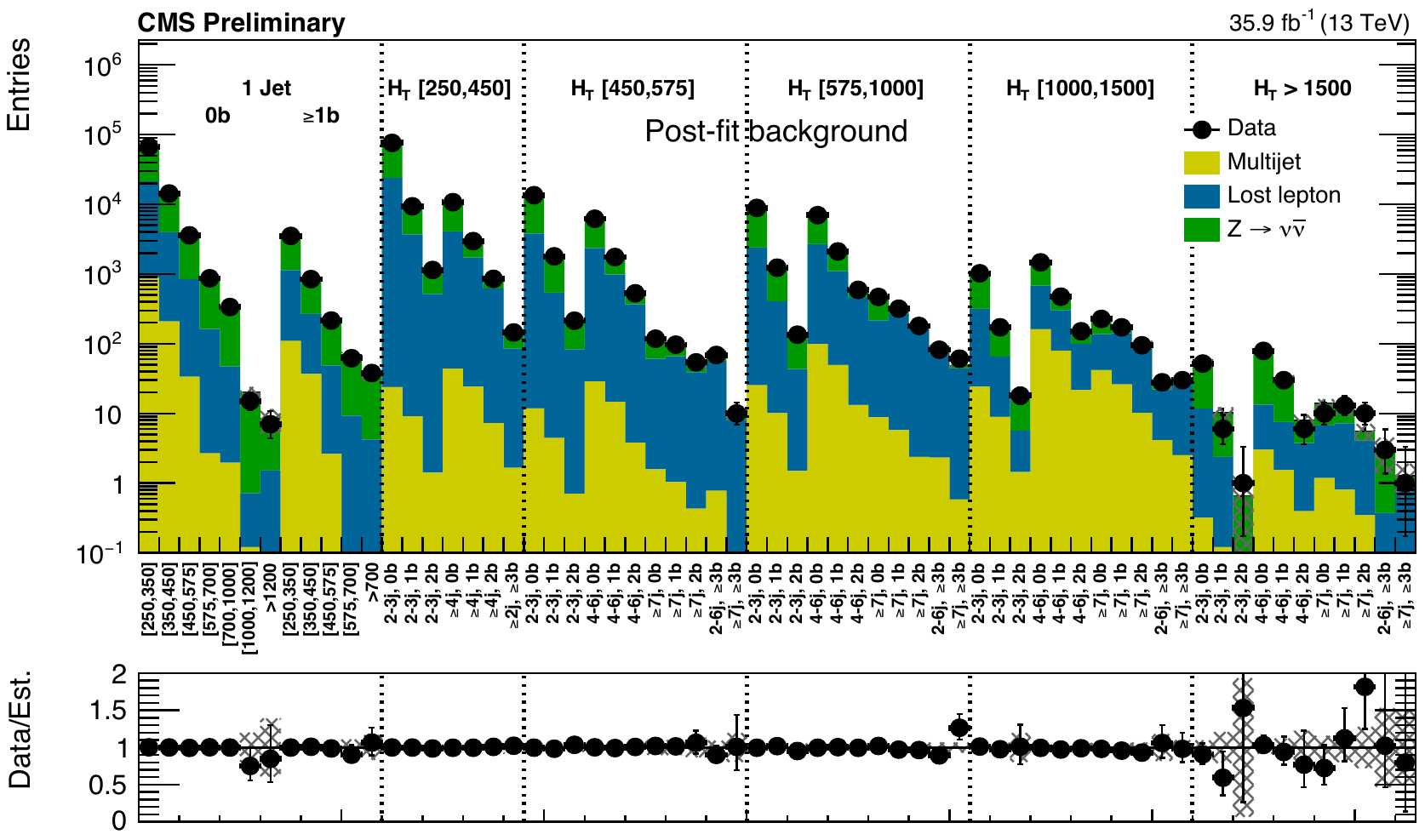}
    \caption{Comparison of post-fit background prediction and observed data events in each topological region.  Hatched bands represent the post-fit uncertainty in the background prediction.
For the monojet, on the $x$-axis the \ptj binning is shown in units of \GeV,
      whereas for the multijet signal regions, the notations j, b indicate \njets, \nbtags labeling.
}
    \label{fig:otherResults_post0}
\end{figure*}

\begin{figure*}[htb]
  \centering
    \includegraphics[width=0.72\textwidth]{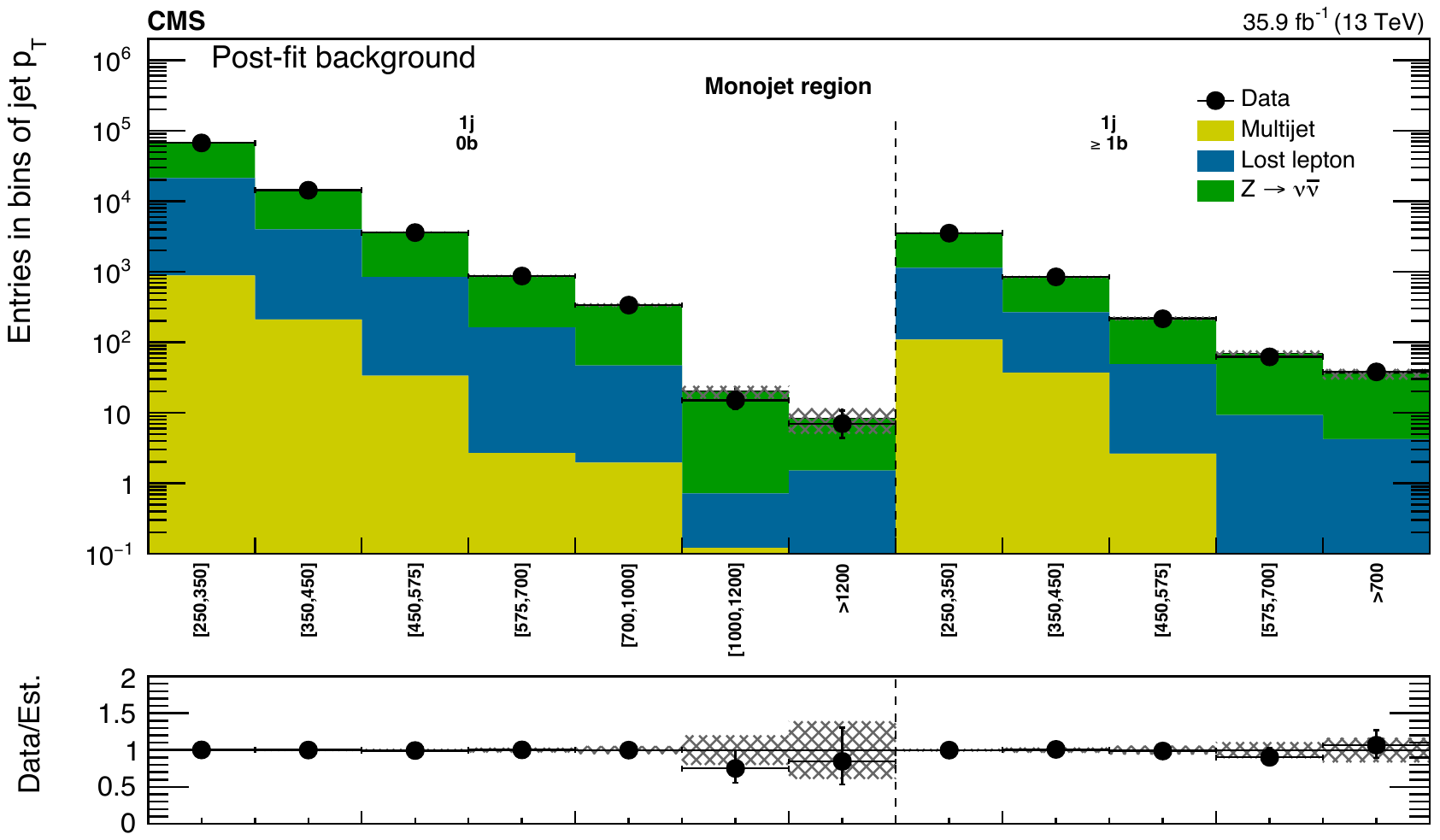}\\
    \includegraphics[width=0.72\textwidth]{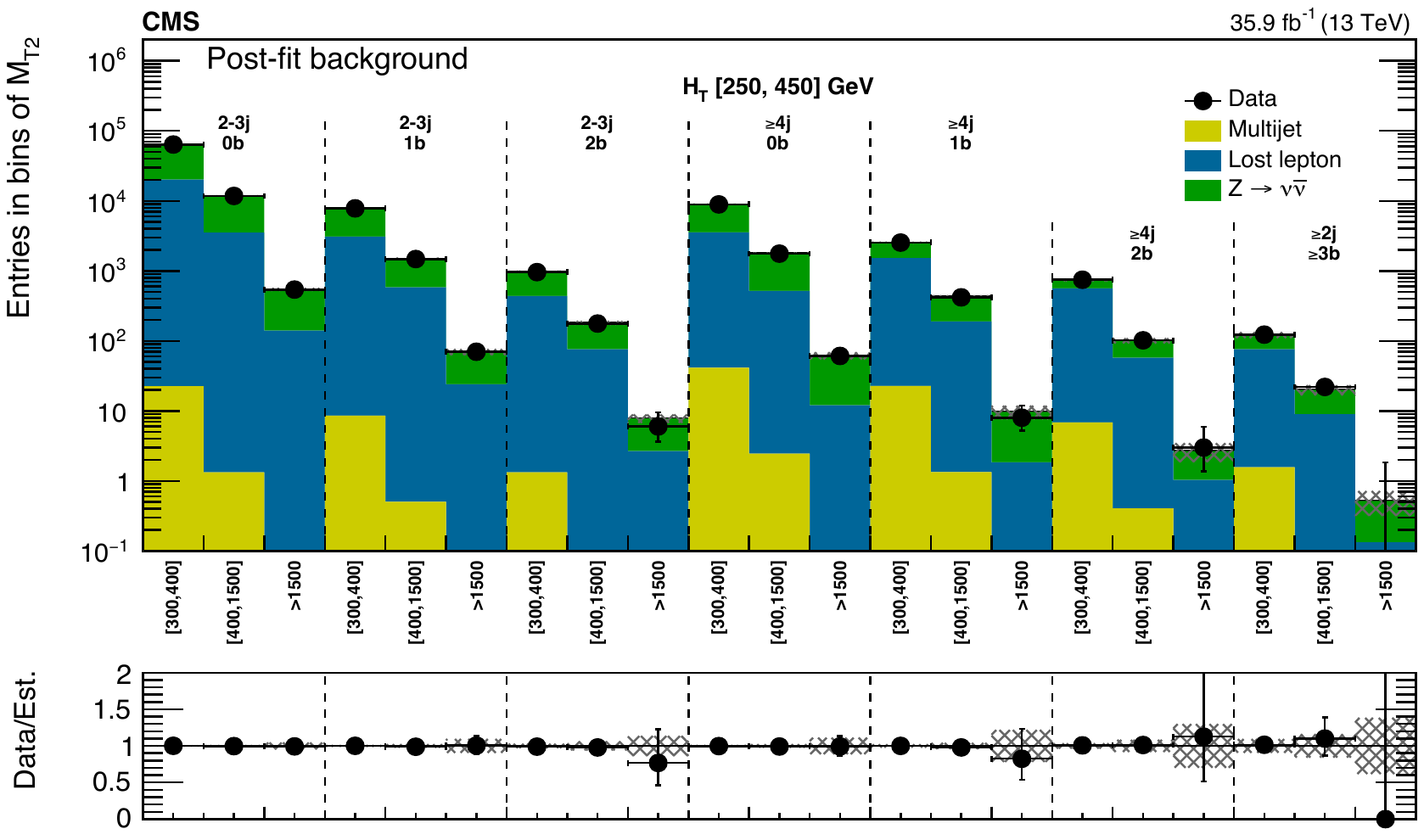}\\
    \includegraphics[width=0.72\textwidth]{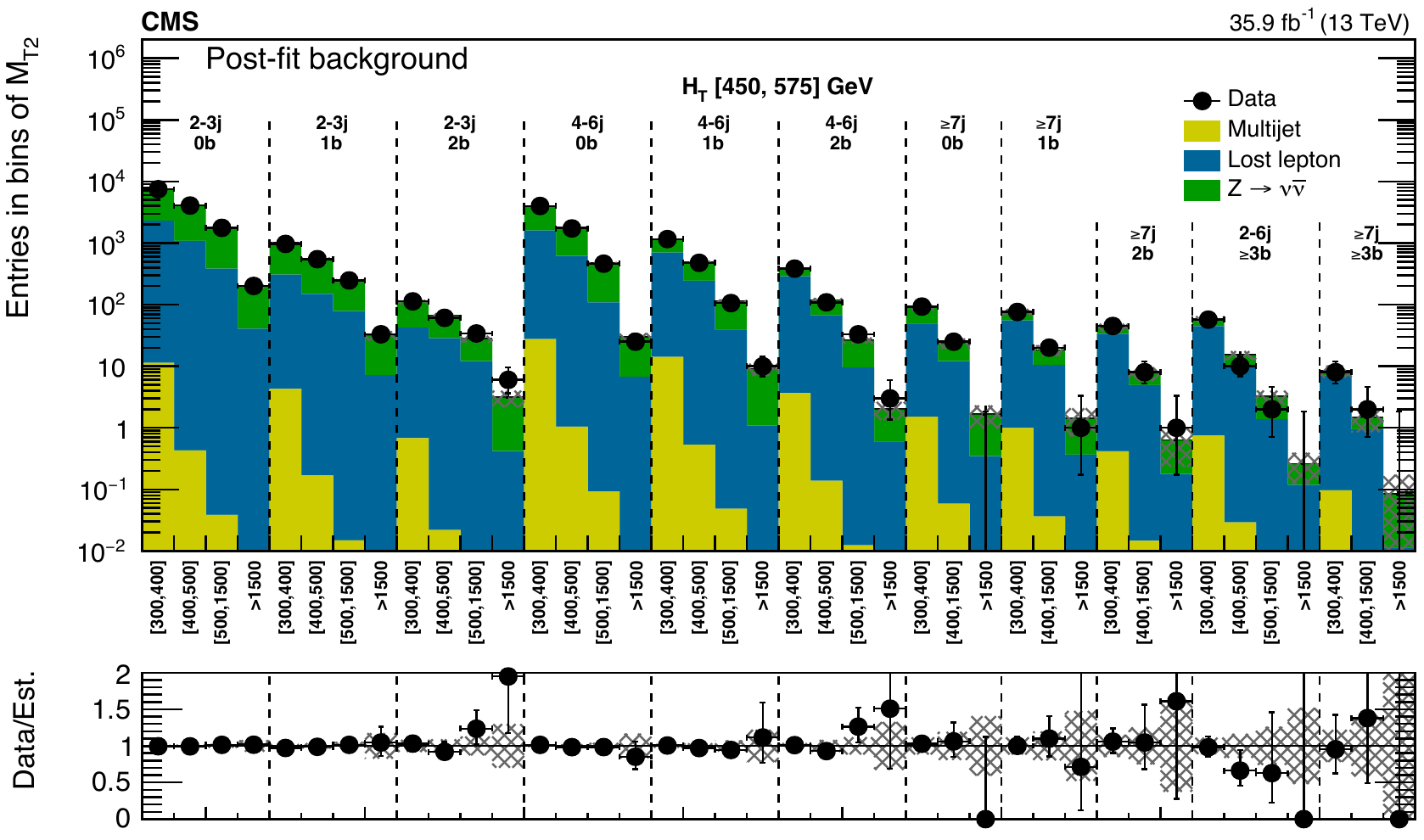}\\
    \caption{(Upper) Comparison of the post-fit background prediction and observed data events in each signal bin in the monojet region. On the $x$-axis, the \ptj binning is shown in units of \GeV.
    (Middle) and (lower): Same for the very low and low-\Ht region.  On the $x$-axis, the \mttwo binning is shown in units of \GeV.
    The hatched bands represent the post-fit uncertainty in the background prediction.}
    \label{fig:otherResults_post1}
\end{figure*}

\begin{figure*}[htb]
  \centering
    \includegraphics[width=0.72\textwidth]{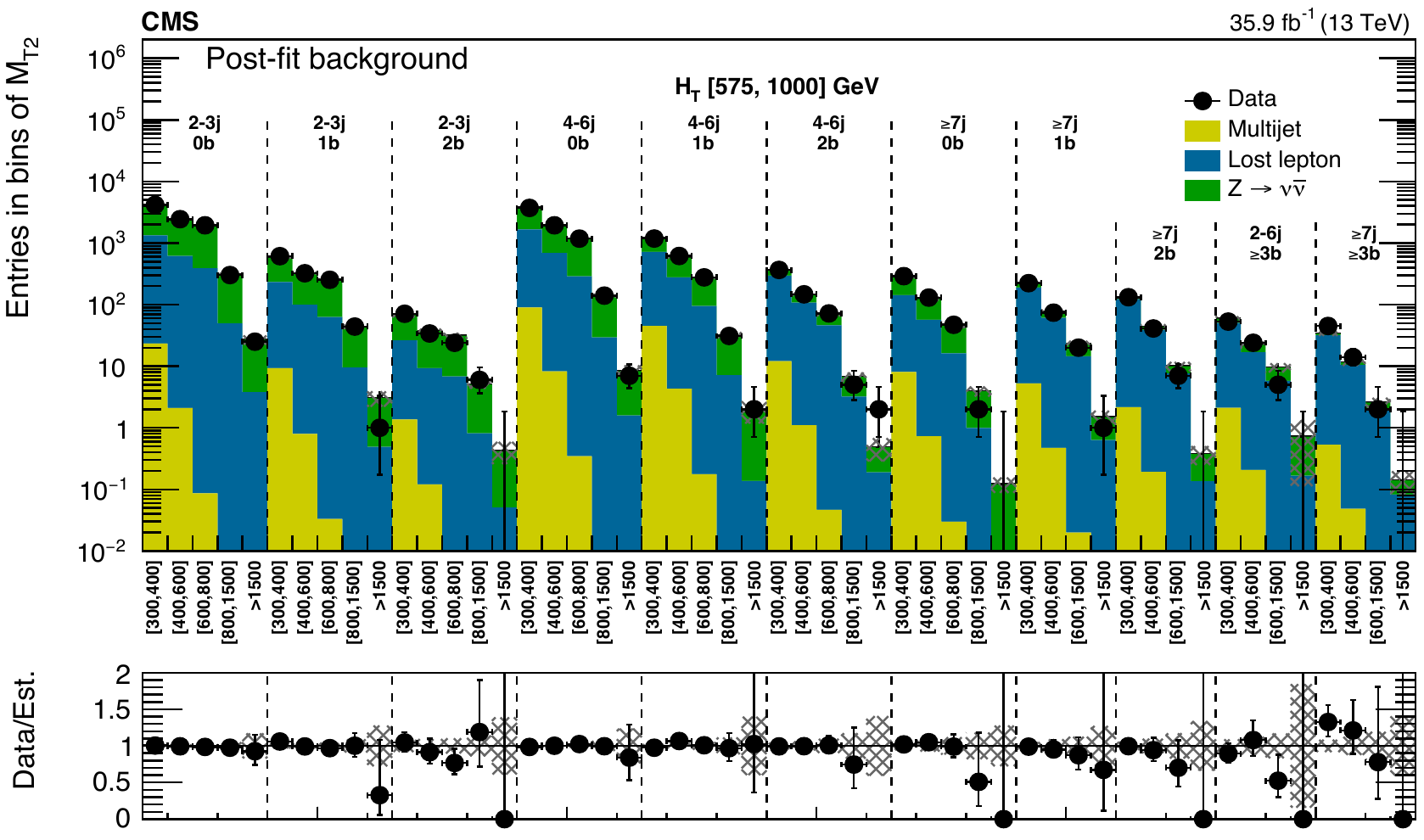}\\
    \includegraphics[width=0.72\textwidth]{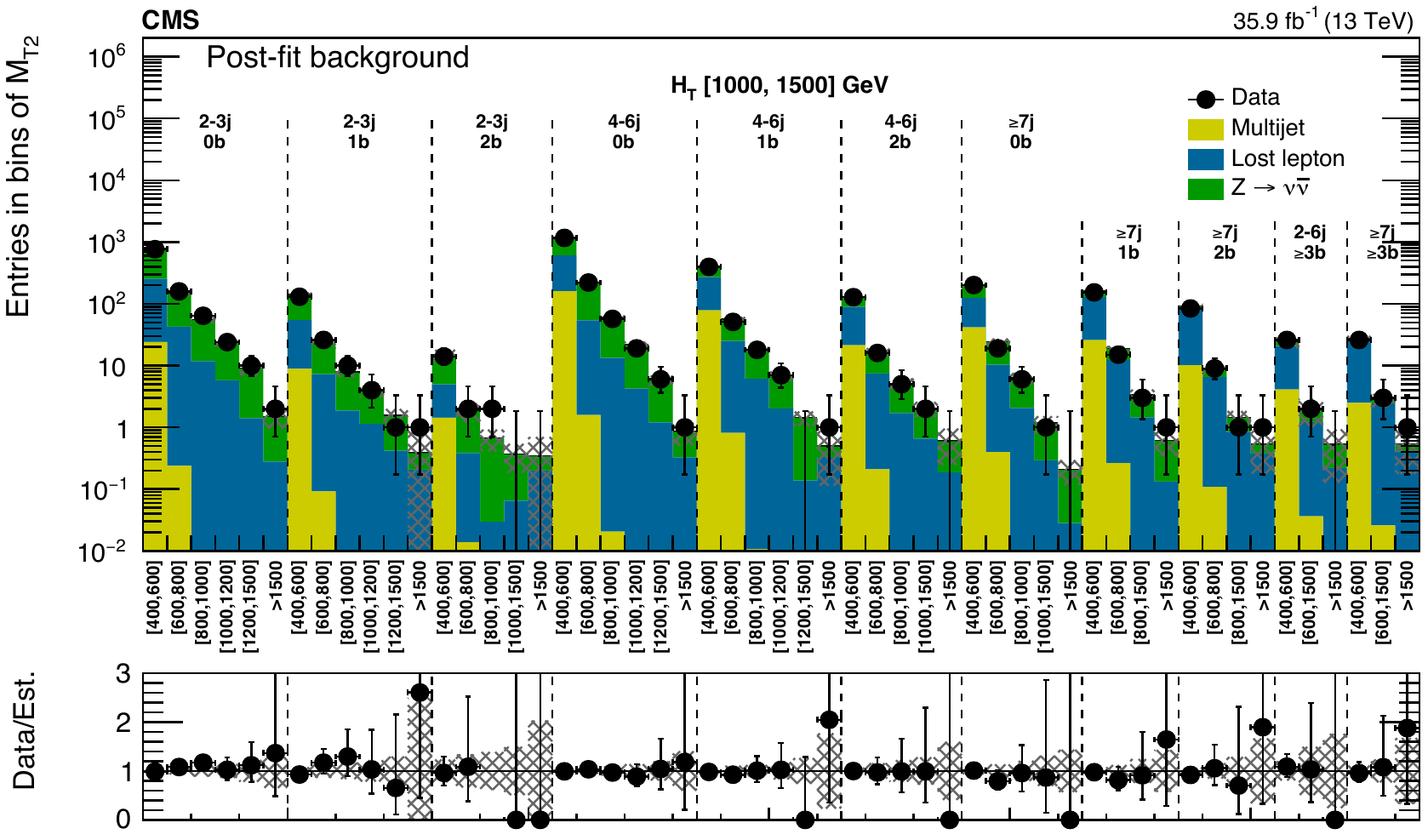}\\
    \includegraphics[width=0.72\textwidth]{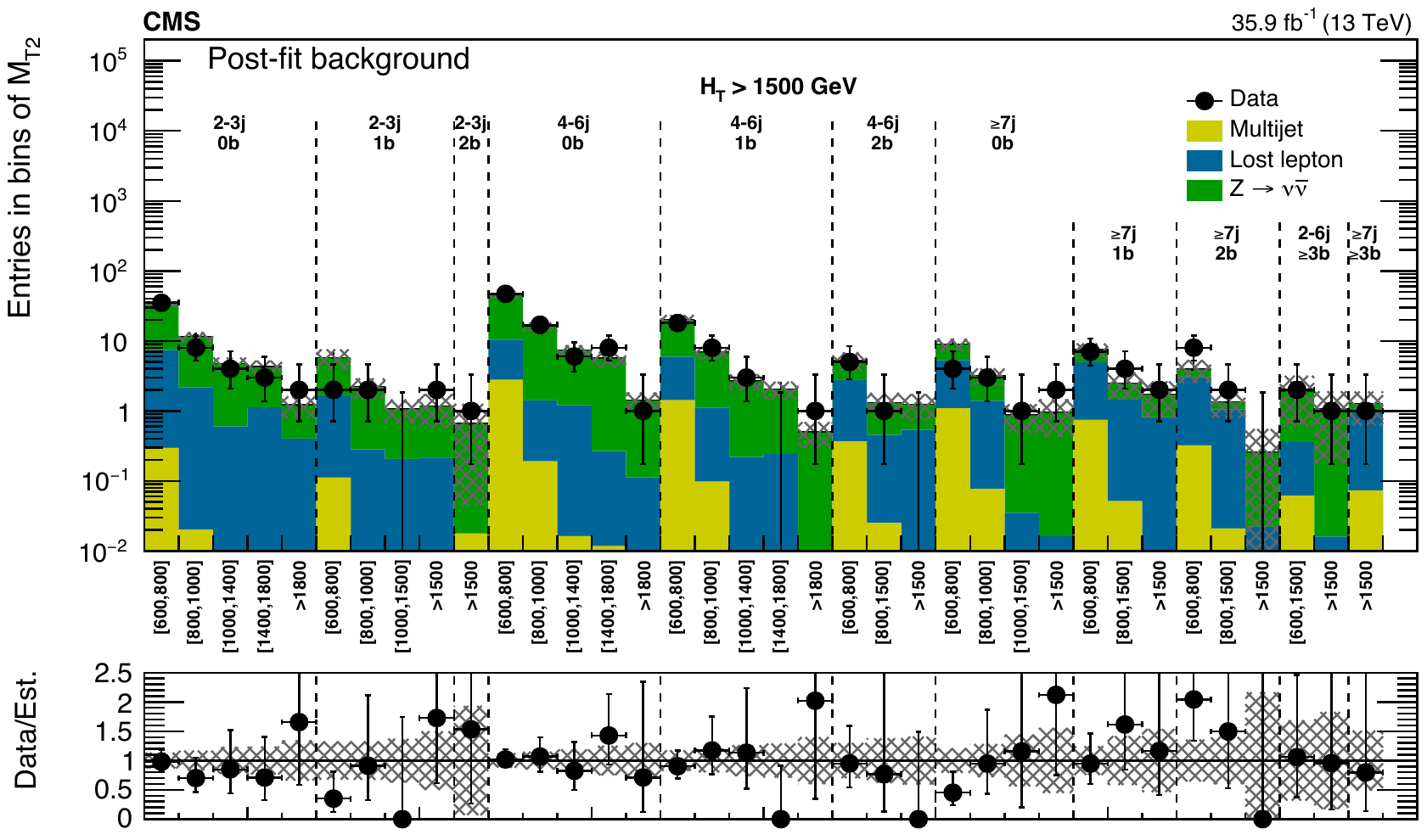}\\
    \caption{(Upper) Comparison of the post-fit background prediction and observed data events in each signal bin in the medium-\Ht region.
    Same for the high- (middle) and extreme- (lower) \Ht regions. On the $x$-axis, the \mttwo binning is shown in units of \GeV.
    The hatched bands represent the post-fit uncertainty in the background prediction.
    For the extreme-\Ht region, the last bin is left empty for visualization purposes.}
    \label{fig:otherResults_post2}
\end{figure*}

\begin{figure*}[htb]
  \centering
    \includegraphics[width=0.79\textwidth]{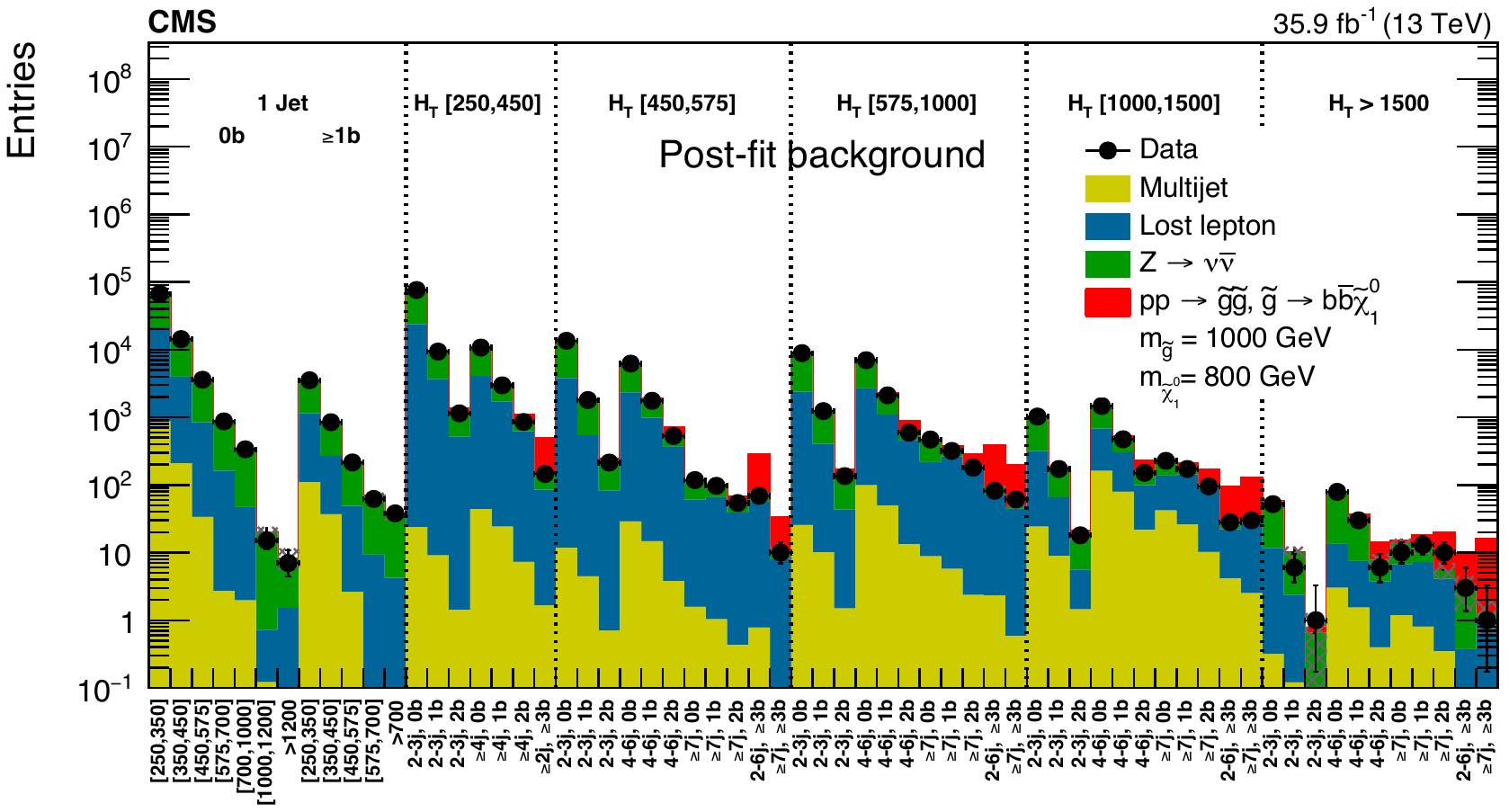}\\
    \includegraphics[width=0.79\textwidth]{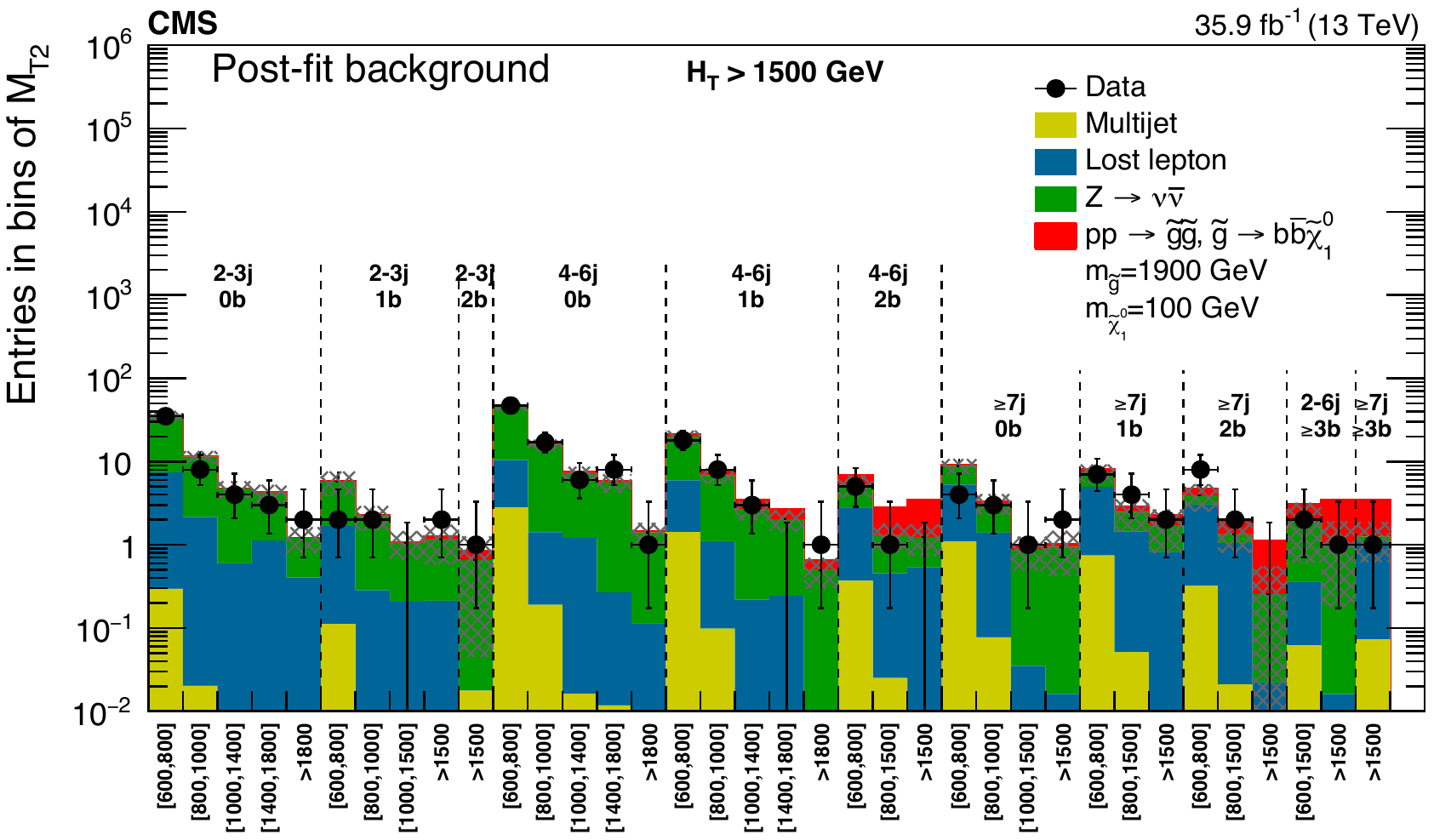}\\
    \caption{(Upper) The post-fit background prediction and observed data events in the analysis binning, for all topological regions with the expected yield for the signal model of gluino mediated bottom-squark production ($m_{\PSg}=1000$\GeV, $m_{\PSGczDo}=800$\GeV) stacked on top of the expected background.
    For the monojet regions, the \ptj binning is in units of \GeV.
    (Lower) Same for the extreme-\Ht region for the same signal with ($m_{\PSg}=1900$\GeV, $m_{\PSGczDo}=100$\GeV).
    On the $x$-axis, the \mttwo binning is shown in units of \GeV.
    The hatched bands represent the post-fit uncertainty in the background prediction.
    For the extreme-\Ht region, the last bin is left empty for visualization purposes.}
    \label{fig:otherResults_post_withSignal}
\end{figure*}

\cleardoublepage \section{The CMS Collaboration \label{app:collab}}\begin{sloppypar}\hyphenpenalty=5000\widowpenalty=500\clubpenalty=5000\input{SUS-16-036-authorlist.tex}\end{sloppypar}
\end{document}

%% file: SUS-16-036-authorlist.tex
\textbf{Yerevan Physics Institute,  Yerevan,  Armenia}\\*[0pt]
A.M.~Sirunyan, A.~Tumasyan
\vskip\cmsinstskip
\textbf{Institut f\"{u}r Hochenergiephysik,  Wien,  Austria}\\*[0pt]
W.~Adam, F.~Ambrogi, E.~Asilar, T.~Bergauer, J.~Brandstetter, E.~Brondolin, M.~Dragicevic, J.~Er\"{o}, M.~Flechl, M.~Friedl, R.~Fr\"{u}hwirth\cmsAuthorMark{1}, V.M.~Ghete, J.~Grossmann, J.~Hrubec, M.~Jeitler\cmsAuthorMark{1}, A.~K\"{o}nig, N.~Krammer, I.~Kr\"{a}tschmer, D.~Liko, T.~Madlener, I.~Mikulec, E.~Pree, D.~Rabady, N.~Rad, H.~Rohringer, J.~Schieck\cmsAuthorMark{1}, R.~Sch\"{o}fbeck, M.~Spanring, D.~Spitzbart, J.~Strauss, W.~Waltenberger, J.~Wittmann, C.-E.~Wulz\cmsAuthorMark{1}, M.~Zarucki
% ...
\vskip\cmsinstskip
% ...
1:~~Also at Vienna University of Technology, Vienna, Austria\\
% ...
